\documentclass[
amsmath,
prd,nofootinbib,floatfix,11pt,
]{revtex4}

\usepackage{placeins}
\usepackage{afterpage}

\usepackage{axodraw2}
\usepackage{graphicx}
\usepackage{setspace}
\usepackage{bm}

\newcommand\beq{\begin{eqnarray}}
\newcommand\eeq{\end{eqnarray}}
\def\lsim{\mathrel{\rlap{\lower4pt\hbox{$\sim$}}
    \raise1pt\hbox{$<$}}}                
\def\gsim{\mathrel{\rlap{\lower4pt\hbox{$\sim$}}
    \raise1pt\hbox{$>$}}}            

\allowdisplaybreaks
\newcommand\MSbar{$\overline{\rm{MS}}$ }
\newcommand\poly{C}

\def\partials{\frac{\partial}{\partial s}}
\def\sqrts{\sqrt{s}}

\def\sqrtw{\sqrt{w}}
\def\sqrtx{\sqrt{x}}
\def\sqrty{\sqrt{y}}
\def\sqrtz{\sqrt{z}}
\begin{document}

\renewcommand{\theequation}{\arabic{section}.\arabic{equation}}
\renewcommand{\thefigure}{\arabic{section}.\arabic{figure}}
\renewcommand{\thetable}{\arabic{section}.\arabic{table}}

\title{\Large \baselineskip=20pt 
Evaluation of three-loop self-energy master integrals with\\ 
four or five propagators}
\author{Stephen P.~Martin}
\affiliation{\mbox{\it Department of Physics, Northern Illinois University, DeKalb IL 60115}}

\begin{abstract}\normalsize \baselineskip=15pt 
I obtain identities satisfied by the 3-loop self-energy master integrals with four or five propagators with generic masses, including the derivatives with respect to each of the squared masses and the external momentum invariant. These identities are then recast in terms of the corresponding renormalized master integrals, enabling straightforward numerical evaluation of them by the differential equations approach. Some benchmark examples are provided. The method used to obtain the derivative identities relies only on the general form implied by integration by parts relations, without actually following the usual integration by parts reduction procedure. As a byproduct, I find a simple formula giving the expansion of the master integrals to arbitrary order in the external momentum invariant, in terms of known derivatives of the corresponding vacuum integrals. 
\end{abstract}

\maketitle

\vspace{-0.2in}

\baselineskip=14.5pt

\tableofcontents

\baselineskip=14.7pt
\newpage

\section{Introduction\label{sec:intro}}
\setcounter{equation}{0}
\setcounter{figure}{0}
\setcounter{table}{0} 
\setcounter{footnote}{1}

In modern evaluations of dimensionally regularized \cite{Bollini:1972ui,Bollini:1972bi,Ashmore:1972uj,Cicuta:1972jf,tHooft:1972tcz,tHooft:1973mfk,Bardeen:1978yd,Braaten:1981dv} loop integrals for quantum field theory, the integration by parts (IBP) relations \cite{Tkachov:1981wb,Chetyrkin:1981qh} often play an important role. By applying IBP relations repeatedly \cite{Laporta:2000dsw,Anastasiou:2004vj,Chetyrkin:2006dh,Faisst:2006sr,Smirnov:2008iw,Smirnov:2014hma,Smirnov:2019qkx,Studerus:2009ye,Gluza:2010ws,vonManteuffel:2012np,Lee:2012cn,Lee:2013mka,vonManteuffel:2014ixa,Larsen:2015ped,Georgoudis:2016wff,Maierhofer:2017gsa,Maierhofer:2018gpa,Klappert:2020nbg}, one can discover identities between different loop integrals with common topological features, allowing one to eliminate many of them in favor of a finite \cite{Smirnov:2010hn} number of master integrals. In particular, derivatives of the master integrals with respect to the propagator squared masses, and with respect to external momentum invariants, can always be written as linear combinations of the master integrals. This results in differential equations whose solution (either analytical or numerical) for the master integrals can be obtained.

The proximate motivation for the present paper was the problem of evaluating self-energy integrals at up to three-loop order for use in the Standard Model, with the eventual goal, certainly not realized in this paper, of evaluating the complete three-loop corrections to the pole masses of the electroweak bosons.
This involves reduction of a general 3-loop self-energy to master integrals, and then the evaluation of the master integrals, using differential equations in the external momentum invariant. In the following, the differential equations satisfied by the 3-loop self-energy master integrals with 4 and 5 propagators will be found explicitly, enabling their numerical computation. For the Standard Model, there are only four distinct large masses, that of the top quark, Higgs boson, and $W$ and $Z$ bosons, so only a subset of the general kinematic 3-loop topologies will be necessary. However, it is useful to have methods that work for general masses, for possible future applications to extensions such as models with supersymmetric particles or new vectorlike quarks and leptons, and other models that may not be forseen at present. The discussion and results below are therefore formulated for generic three-loop self-energy integrals, and it is hoped that some of the ideas may have even broader applicability beyond self-energy integrals. 

In some cases, the reduction to master integrals using IBP identities  can be challenging, due to their number and complexity. In this paper, I will employ a different method, which makes use of the general form for results implied by the IBP relations, without actually using the IBP reduction procedure itself. The idea will be described in terms of self-energy integrals involving an external momentum 
$p^\mu$, in
\beq
d = 4 - 2\epsilon
\eeq
dimensions, assumed to be either Euclideanized or to have the metric signature with mostly $+$ signs, so that the external momentum invariant is
\beq
s = -p^2.
\eeq
The integrals also depend on some number of internal propagator squared masses denoted $x,y,\ldots$. The IBP procedure leads to identities that can always be written in the form:
\beq
\sum_k \poly_k\,  {\bf I}_k(s;x,y,\ldots) &=& 0
,
\label{eq:genericidentity}
\eeq
where the ${\bf I}_k(s;x,y,\ldots$ are the loop integrals, and it is a crucial feature that the $\poly_k$ are polynomials in $s$, in the internal squared masses, and in $\epsilon$.

The idea to be exploited here is to obtain the identities of the form eq.~(\ref{eq:genericidentity}), not by repeatedly applying IBP relations, but by making a guess for the degree in $s$ of each of the polynomials, and writing the most general form for each polynomial $\poly_k$ in terms of a finite number of unknown coefficients. Then, after expanding the loop integrals ${\bf I}_k(s;x,y,\ldots)$ in small $s$, the unknown polynomial coefficients in the $\poly_k$ can be fixed by requiring each power of $s$ in the expansion of eq.~(\ref{eq:genericidentity}) to have vanishing coefficient. If the degree in $s$ of any one of the polynomials $\poly_k$ has been incorrectly guessed to be too low, this procedure will encounter a contradiction. If the guessed degrees in $s$ are minimal, one may obtain a unique solution for the unknown coefficients after expanding eq.~(\ref{eq:genericidentity}) in $s$ to some finite power, after which the next few powers in $s$ will give consistency checks. If the guessed degree in $s$ for one or more of the polynomials is larger, then one will find multi-parameter consistent solutions for the polynomial coefficients, which can be resolved by setting any unnecessary coefficients (of the highest powers of $s$ in the $C_k$) to 0. 

Of course, this method relies on the ability to evaluate the expansions in $s$ of the integrals ${\bf I}_k$ to sufficiently high order. That is particularly straightforward for the examples described below, which are the 3-loop self-energy integrals with 4 or 5 propagators with arbitrary squared masses. In this paper, I will find the master integrals and identities relating them, including the results needed to numerically evaluate them using the differential equations approach \cite{Kotikov:1990kg,Kotikov:1991hm,Ford:1991hw,Ford:1992pn,Bern:1993kr,Remiddi:1997ny,Caffo:1998du,Gehrmann:1999as,Caffo:2002ch,Caffo:2002wm,Caffo:2003ma,Martin:2003qz,Martin:2005qm,Martin:2016bgz,Martin:2021pnd}.

Note that the method used here works even if the small $s$ expansions for the integrals fail to converge for realistic physical values. The method has several other advantages. First, because one is looking for a finite set of integer polynomial coefficients, one can find them by assigning arbitrary rational numbers to all of the squared masses $x,y,\ldots$ and even to $\epsilon$, then repeating the process with different rational numbers until either all coefficients have been successfully identified, or until a contradiction has been encountered. (In the latter case, one increases the degrees of the polynomials, and tries again.) That was the method used to obtain the results below; it greatly reduces the computer memory and processing requirements, making the calculation tractable in cases where it might be much more difficult otherwise.  
The use of rational numbers is similar to strategies described in the recent literature for using finite fields and rational fields  to reconstruct identities between integrals, which follow from early work in ref.~\cite{vonManteuffel:2014ixa} and \cite{Peraro:2019svx}. Several public codes employ these methods, including FiniteFlow \cite{Peraro:2019svx}, the FIRE6 \cite{Smirnov:2019qkx} IBP code, and FireFly \cite{Klappert:2019emp,Klappert:2020aqs} and the Kira 2.0 \cite{Klappert:2020nbg} IBP code, and Caravel \cite{Abreu:2020xvt} based on numerical unitarity. 

A second advantage is that when one is evaluating a physical observable, one need not solve for all of the individual reducible integrals that may appear in it, or for other  reducible integrals in the same sectors, which are often vast in number. Instead, one can choose one of the ${\bf I}_k$ to be the whole integral expression (typically including irreducible numerator factors) for the contribution to the observable in question with a given diagram topology, and let the others be the master integrals, which will have been previously identified by finding other identities that eliminate all other candidate masters. If the small $s$ expansion of the observable can be obtained, it can be used to find the required polynomial coefficients expressing it in terms of master integrals, again even if the expansion fails to converge for the physical values of $s$ and other parameters. A third advantage is that it allows one to confidently make statements such as ``no identity relating the following integrals exists, for polynomials $\poly_k$ up to degrees $n_k$ in $s$". Such statements are harder to be completely certain of using only the IBP procedure, since there are an infinite number of IBP relations, and it is not even guaranteed that the IBP relations capture all possible valid identities between integrals. 

One slight disadvantage must be admitted: one cannot be absolutely certain (in the sense of a rigorous mathematical proof) that an identity that one has obtained is correct, since it could be that some contradiction will be encountered after the expansion in $s$ has been extended beyond the particular level that one has chosen. However, rigorous proofs aside, it seems extremely unlikely that an incorrect identity would survive checks if the expansion in $s$ has been extended several levels beyond that necessary to uniquely fix all of the unknown coefficients. Remaining doubts can be reduced to an infinitesimal level by simply further extending the expansion in $s$.

It should also be noted that the expansion need not be in small $s$; for example, one could instead expand in some or all of the squared masses treated as small. One could also use a large $s$ expansion to solve for the polynomial coefficients, or even combine constraints on the polynomial coefficients obtained from different expansions. The small $s$ expansion was chosen here because of the convenient availability \cite{Martin:2016bgz} of arbitrary derivatives of vacuum (no external momenta) master integrals through 3-loop order. A somewhat similar proposal, based on a still different type of expansion, may be found in ref.~\cite{Liu:2018dmc}, and another approach for obtaining identities while avoiding the use of huge numbers of IBP relations can be found in refs.~\cite{Mastrolia:2018uzb,Frellesvig:2019kgj,Frellesvig:2019uqt}.

The rest of this paper is organized as follows. In section \ref{sec:notations}, I give my notations and conventions for the relevant scalar loop integrals without numerators, which adhere to those used in refs.~\cite{Martin:2003qz,Martin:2005qm,Martin:2016bgz,Martin:2021pnd}. Section \ref{sec:expansions} gives a simple formula for the expansion to arbitrary order in small $s$ for a large class of self-energy integrals, including all of the ones discussed in this paper, in terms of known derivatives of vacuum master integrals. Sections \ref{sec:fourprop} and \ref{sec:fiveprop} provide the identities for the 3-loop master integrals with four and five propagators, respectively. Other useful approaches to calculating 3-loop vacuum and self-energy integrals are found in refs.~\cite{Broadhurst:1991fi,Berends:1993ee,Avdeev:1994db,Fleischer:1994dc,Avdeev:1995eu,Broadhurst:1998rz,Fleischer:1999mp,Steinhauser:2000ry,Schroder:2005va,Davydychev:2003mv,Kalmykov:2005hb,Kalmykov:2006pu,Bytev:2009mn,Bekavac:2009gz,Bytev:2009kb,Bytev:2011ks,Grigo:2012ji,Bauberger:2017nct,Bauberger:2019heh,Dubovyk:2021lqe,Dubovyk:2022frj,Broadhurst:2022bkw,ReyesR:2022ssk}.
In section \ref{sec:numerical}, I describe the numerical computation of the master integrals using the differential equations method, and give some benchmark values. Section \ref{sec:outlook} has some concluding remarks.

\section{Notations and conventions\label{sec:notations}}
\setcounter{equation}{0}
\setcounter{figure}{0}
\setcounter{table}{0} 
\setcounter{footnote}{1}

\begin{figure}[!t]
\begin{center}
\includegraphics[width=0.9\linewidth,angle=0]{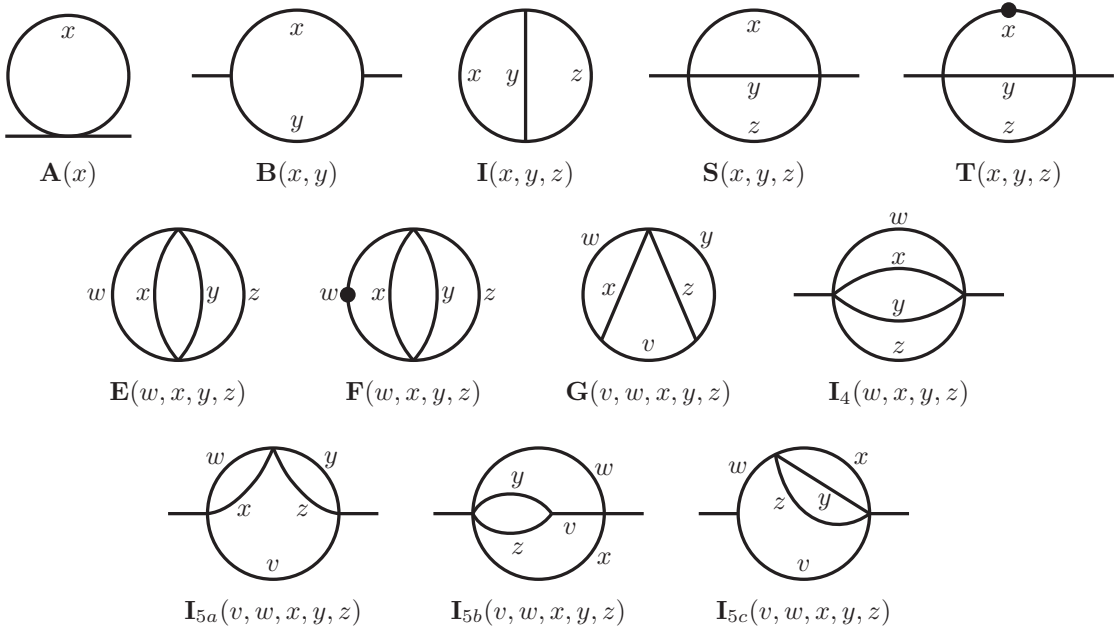}
\end{center}
\vspace{-0.4cm}
\begin{minipage}[]{0.95\linewidth}
\caption{\label{fig:masters}
Diagrams for vacuum and self-energy integrals appearing in this paper, as defined in 
eqs.~(\ref{eq:defboldA})-(\ref{eq:defboldI5c}), following the same conventions and notations used in refs.~\cite{Martin:2003qz,Martin:2005qm,Martin:2016bgz,Martin:2021pnd}. The labels $v,w,x,y,z$ on the internal lines denote the propagator squared masses.}
\end{minipage}
\end{figure}

In the following, consider loop momentum integrals in $d = 4 - 2 \epsilon$ Euclidean dimensions, written
in terms of 
\beq
\int_k &\equiv& 16 \pi^2 \mu^{2 \epsilon} \int \frac{d^d k}{(2 \pi)^d}.
\eeq
The integrals appearing in this paper are shown in Figure \ref{fig:masters}.
The 1-loop vacuum and self-energy master integrals are 
\beq
{\bf A}(x) &=& \int_k \frac{1}{k^2 + x}
\,=\,
x \left (\frac{4 \pi \mu^2}{x}\right )^\epsilon \Gamma(\epsilon - 1),
\label{eq:defboldA}
\\
{\bf B}(x,y) &=& \int_k \frac{1}{[k^2 + x][(k-p)^2 + y]},
\eeq
and at two loops,
\beq
{\bf I}(x,y,z) &=& \int_k\int_q \frac{1}{[k^2 + x][q^2 + y][(k+q)^2 + z]},
\\
{\bf S}(x,y,z) &=& \int_k\int_q \frac{1}{[k^2 + x][q^2 + y][(k+q-p)^2 + z]},
\\
{\bf T}(x,y,z) &=& \int_k\int_q \frac{1}{[k^2 + x]^2\,[q^2 + y][(k+q-p)^2 + z]}
.
\eeq
The three-loop vacuum and self-energy masters are denoted by
\beq
{\bf E}(w,x,y,z) &=& \int_k\int_q\int_r \frac{1}{[k^2 + w][q^2 + x][r^2 + y][(k+q+r)^2 + z]}
,
\\
{\bf F}(w,x,y,z) &=& \int_k\int_q\int_r \frac{1}{[k^2 + w]^2 \,[q^2 + x][r^2 + y][(k+q+r)^2 + z]}
,
\\
{\bf G}(v,w,x,y,z) &=& \int_k\int_q\int_r 
\frac{1}{[k^2 + v][q^2 + w][(k+q)^2 + x][r^2 + y][(k+r)^2 + z]}
,
\\
{\bf I}_4(w,x,y,z) &=& \int_k\int_q\int_r \frac{1}{[k^2 + w][q^2 + x][r^2 + y][(k+q+r-p)^2 + z]}
,
\\
{\bf I}_{5a}(v,w,x,y,z) &=& \int_k\int_q\int_r 
\frac{1}{[k^2 + v][q^2 + w][(k+q-p)^2 + x][r^2 + y][(k+r-p)^2 + z]}
,
\phantom{xxx}
\\
{\bf I}_{5b}(v,w,x,y,z) &=& \int_k\int_q\int_r 
\frac{1}{[k^2 + v][q^2 + w][(k+q-p)^2 + x][r^2 + y][(k+r)^2 + z]}
,
\phantom{xxx}
\\
{\bf I}_{5c}(v,w,x,y,z) &=& \int_k\int_q\int_r 
\frac{1}{[k^2 + v][(k-p)^2 + w][q^2 + x][r^2 + y][(k+q+r-p)^2 + z]}
.
\phantom{xxx}
\label{eq:defboldI5c}
\eeq
Note that the external momentum invariant $s$ is omitted from the arguments of the self-energy
integral functions. The integral functions defined above have various symmetries under interchange
of the squared mass arguments, which are obvious from the diagrams in Figure \ref{fig:masters},
and will be used below without commentary.
The integral ${\bf E}(w,x,y,z)$ is sometimes convenient because of its symmetry properties, but it is technically not a master integral because it can be eliminated in favor of the ${\bf F}$ integrals, through the identity
\beq
(3 \epsilon - 2) {\bf E}(w,x,y,z) = 
w {\bf F}(w,x,y,z) +
x {\bf F}(x,w,y,z) +
y {\bf F}(y,w,x,z) +
z {\bf F}(z,w,x,y) ,
\label{eq:EEFFidentity}
\eeq
which follows from dimensional analysis.

In the following, we will use two different notations for derivatives with respect to a squared
mass $x$, depending on the typographical situation. In some cases, we will write $\partial_x$, while in other cases we will use a prime on a squared mass argument of a function to denote differentiation with respect to that argument. For example,
\vspace{-10pt}
\beq
{\bf T}(x,y,z) \,=\, -\partial_x {\bf S}(x,y,z) \,=\, -{\bf S}(x',y,z) ,
\eeq
%
and
\vspace{-10pt}
\beq
{\bf F}(w,x,y,z) \,=\, -\partial_w {\bf E}(w,x,y,z) \,=\, -{\bf E}(w',x,y,z) ,
\label{eq:defineboldF}
\eeq
and for a generic function,
\beq
f(x',y,x'') &=& \partial_x \partial^2_z f(x,y,z) \Bigl |_{z = x}.
\eeq

It is convenient to write expressions for physical observables in terms of renormalized master integrals, which are obtained from the above by subtracting ultraviolet (UV) divergences in a particular way, then taking the limit $\epsilon \rightarrow 0$, and writing the results in terms of the scale $Q$ defined by 
\beq
Q^2 = 4\pi e^{-\gamma} \mu^2.
\eeq
If the \MSbar renormalization scheme \cite{Bardeen:1978yd,Braaten:1981dv} is used, then $Q$ is the renormalization scale. (This does not obligate one to use the \MSbar scheme, however.) 

As explained in ref.~\cite{Martin:2021pnd}, the renormalized master integrals have the key advantage that expansions
of the master integrals at a given loop order to positive powers of $\epsilon$ are never needed, even for calculations
at higher loop order. (In fact, in practice this feature provides a very useful consistency check on calculations.) The renormalized $\epsilon$-finite basis of master integrals thus constitute an optimal and minimal set for expressing physical results. In general, this assumes that 
one has first chosen an $\epsilon$-finite basis, in the sense of Chetyrkin, Faisst, Sturm, and Tentyukov in ref.~\cite{Chetyrkin:2006dh}, who showed that it is always possible to find a basis such that the coefficients multiplying the master integrals in an arbitrary observable are finite as $\epsilon \rightarrow 0$. In the present paper, since the masses are treated as generic, this is trivial; any basis defined in terms of basic integrals is $\epsilon$-finite (unless one introduces poles in $\epsilon$ by hand). For special cases in which masses either vanish or are equal to each other or are at thresholds, one should first identify (or verify) the $\epsilon$-finite basis using the algorithm of ref.~\cite{Chetyrkin:2006dh} or by other means, then renormalize the integrals as described below. For more details, and explicit examples at up to three loop order, of the feature that renormalized $\epsilon$-finite master integrals indeed do not require evaluation of the components of positive powers in the expansions in $\epsilon$, see refs.~\cite{Ford:1992pn,Martin:2001vx,Martin:2005eg,Martin:2005ch,Martin:2014cxa,Martin:2015lxa,Martin:2015rea,Martin:2016xsp,Martin:2017lqn,Martin:2022qiv}. At least in the case of ref.~\cite{Martin:2022qiv}, the presence of infrared divergences in $\epsilon$ in individual diagrams does not cause problems; in the other papers listed, infrared divergences were dealt with instead by including infinitesimal masses, but I believe this is not necessary.

Each renormalized integral is denoted by a non-boldfaced letter corresponding to the boldfaced letters in the definitions above, and includes counterterms for each ultraviolet-divergent subdiagram. Explicitly, one defines
\beq
A(x) &=& \lim_{\epsilon \rightarrow 0} [{\bf A}(x) + x/\epsilon] \>=\> x \ln(x/Q^2) - x,
\label{eq:defineArenorm}
\\
B(x,y) &=& \lim_{\epsilon \rightarrow 0} [{\bf B}(x,y) + 1/\epsilon]
\eeq
at 1-loop order, and 
\beq
S(x,y,z) &=& \lim_{\epsilon \rightarrow 0} \left [ {\bf S}(x,y,z)
- S^{1,\,\rm div}(x,y,z)
- S^{2,\,\rm div}(x,y,z) \right ],
\eeq
where the one-loop and two-loop UV sub-divergences are:
\beq
S^{1,\,\rm div}(x,y,z) &=&
\frac{1}{\epsilon} \left [ {\bf A}(x) + {\bf A}(y) + {\bf A}(z)\right ],
\\
S^{2,\,\rm div}(x,y,z) &=& \frac{1}{2\epsilon^2} (x+y+z) + \frac{1}{2\epsilon}
(s/2 - x - y- z).
\eeq
From this, one also has
\beq
I(x,y,z) &=& S(x,y,z) \Bigl |_{s=0},
\\
T(x,y,z) &=& -S(x',y,z). 
\eeq
For the three-loop self-energy integrals, one defines
\beq
I_X(w,x,y,z) &=& \lim_{\epsilon \rightarrow 0} \Bigl [
{\bf I}_X(w,x,y,z)
- {\bf I}_X^{1,\rm div}(w,x,y,z)
- {\bf I}_X^{2,\rm div}(w,x,y,z)
- {\bf I}_X^{3,\rm div}(w,x,y,z)
\Bigr ]
,
\phantom{xxxx}
\label{eq:defineIXrenorm}
\eeq
for $X = 4$, $5a$, $5b$, and $5c$, where the UV sub-divergences are
\beq
{\bf I}_4^{1,\rm div}(w,x,y,z) &=& 
\frac{1}{\epsilon} 
\Bigl [
{\bf A}(w) {\bf A}(x) 
+
{\bf A}(w) {\bf A}(y)
+{\bf A}(w) {\bf A}(z)
+{\bf A}(x) {\bf A}(y)
+{\bf A}(x) {\bf A}(z)
\nonumber \\ &&
+{\bf A}(y) {\bf A}(z)
\Bigr ]
,
\label{eq:I41div}
\\
{\bf I}_4^{2,\rm div}(w,x,y,z) &=&
\left [ \left (\frac{1}{2\epsilon^2} - \frac{1}{2\epsilon} \right )(x+y+z) 
+ \frac{1}{4 \epsilon} (s+w) \right ]
{\bf A}(w)
\nonumber \\ && 
+ (w \leftrightarrow x)
+ (w \leftrightarrow y)
+ (w \leftrightarrow z)
, \phantom{xxxx}
\label{eq:I42div}
\\
{\bf I}_4^{3,\rm div}(w,x,y,z) &=&
\frac{s^2}{36\epsilon} + \left ( \frac{1}{6 \epsilon^2} - \frac{1}{8\epsilon} \right ) s (w+x+y+z)
+ \left (\frac{1}{6 \epsilon^2} - \frac{3}{8 \epsilon} \right )
(w^2 + x^2 + y^2 + z^2)
\nonumber \\ && 
+ \left ( \frac{1}{3\epsilon^3} - \frac{2}{3\epsilon^2} + \frac{1}{3\epsilon} \right )
(w x + w y + w z + x y + x z + y z)
,
\label{eq:I43div}
\eeq
and
\beq
{\bf I}_{5a}^{1,\rm div}(v,w,x,y,z) &=&
\frac{1}{\epsilon} \bigl [{\bf S}(v,w,x) + {\bf S}(v,y,z) \bigr ]
,
\\
{\bf I}_{5a}^{2,\rm div}(v,w,x,y,z) &=&
-\frac{1}{\epsilon^2} {\bf A}(v)
+\left (\frac{1}{2 \epsilon} - \frac{1}{2 \epsilon^2}\right ) 
\bigl  [{\bf A}(w) + {\bf A}(x)+ {\bf A}(y) + {\bf A}(z)  \bigr ]
,
\\
{\bf I}_{5a}^{3,\rm div}(v,w,x,y,z) &=&
\left ( -\frac{1}{6 \epsilon^2} + \frac{1}{12 \epsilon} \right ) s
+
\left ( -\frac{1}{6 \epsilon^3} + \frac{1}{2 \epsilon^2} - \frac{2}{3\epsilon} \right )
(w+x+y+z)
\nonumber \\ && 
+
\left ( -\frac{1}{3 \epsilon^3} + \frac{1}{3 \epsilon^2} + \frac{1}{3\epsilon} \right ) v
,
\eeq
and
\beq
{\bf I}_{5b}^{1,\rm div}(v,w,x,y,z) &=&
\frac{1}{\epsilon} \bigl [{\bf S}(v,w,x) + {\bf I}(v,y,z) \bigr ]
,
\\
{\bf I}_{5b}^{2,\rm div}(v,w,x,y,z) &=&
-\frac{1}{\epsilon^2} {\bf A}(v) 
+\left (\frac{1}{2 \epsilon} - \frac{1}{2 \epsilon^2}\right ) 
\bigl [{\bf A}(w) + {\bf A}(x)+ {\bf A}(y) + {\bf A}(z)  \bigr ]
,
\\
{\bf I}_{5b}^{3,\rm div}(v,w,x,y,z) &=&
\left ( -\frac{1}{12 \epsilon^2} + \frac{5}{24 \epsilon} \right ) s
+
\left ( -\frac{1}{6 \epsilon^3} + \frac{1}{2 \epsilon^2} - \frac{2}{3\epsilon} \right )
(w+x+y+z)
\nonumber \\ && 
+
\left ( -\frac{1}{3 \epsilon^3} + \frac{1}{3 \epsilon^2} + \frac{1}{3\epsilon} \right ) v
,
\eeq
and
\beq
{\bf I}_{5c}^{1,\rm div}(v,w,x,y,z) &=&
\frac{1}{\epsilon} {\bf B}(v,w) \bigl [{\bf A}(x) + {\bf A}(y) + {\bf A}(z) \bigr ] 
,
\\
{\bf I}_{5c}^{2,\rm div}(v,w,x,y,z) &=&
-\frac{1}{4\epsilon} {\bf A}(v) 
+ \left (\frac{1}{2 \epsilon} - \frac{1}{2 \epsilon^2}\right ) 
  \bigl [{\bf A}(x)+ {\bf A}(y) + {\bf A}(z)  \bigr ]
\nonumber \\ &&
+ \left [\left (\frac{1}{2\epsilon^2} - \frac{1}{2\epsilon} \right ) (x+y+z) + 
\frac{1}{4\epsilon} w \right] {\bf B}(v,w)  
,
\\
{\bf I}_{5c}^{3,\rm div}(v,w,x,y,z) &=&
-\frac{1}{12 \epsilon} s + 
\left (-\frac{1}{6 \epsilon^2} + \frac{3}{8\epsilon}\right ) (v+w)
+
\left ( -\frac{1}{3 \epsilon^3} + \frac{2}{3 \epsilon^2} - \frac{1}{3\epsilon} \right )
(x+y+z)
.
\phantom{xxx}
\eeq
Also, one has
\beq
E(w,x,y,z) &=& I_4(w,x,y,z) \Bigl |_{s=0},
\\
F(w,x,y,z) &=& -I_4(w',x,y,z) \Bigl |_{s=0},
\\
G(v,w,x,y,z) &=& I_{5a}(v,w,x,y,z) \Bigl |_{s=0} \,=\, I_{5b}(v,w,x,y,z) \Bigl |_{s=0},
\eeq
as in ref.~\cite{Martin:2016bgz}.
The renormalized integrals have a dependence on $Q$ given by:
\beq
Q^2 \frac{\partial}{\partial Q^2} A(x) &=& -x,
\\
Q^2 \frac{\partial}{\partial Q^2} B(x,y) &=& 1,
\\
Q^2 \frac{\partial}{\partial Q^2} I(x,y,z) &=& A(x) + A(y) + A(z) - x - y - z,
\\
Q^2 \frac{\partial}{\partial Q^2} S(x,y,z) &=& A(x) + A(y) + A(z) - x - y - z + s/2,
\\
Q^2 \frac{\partial}{\partial Q^2} T(x,y,z) &=& -A(x)/x,
\\
Q^2 \frac{\partial}{\partial Q^2} F(w,x,y,z) &=& 
\Bigl [x+y+z -w-A(x) - A(y) - A(z)\Bigr ] A(w)/w 
+ 7 w/4,
\\
Q^2 \frac{\partial}{\partial Q^2} I_{4}(w,x,y,z) &=& 
2 A(w) A(x) 
+ 2 A(w) A(y) 
+ 2 A(w) A(z) 
\nonumber \\ &&
+ 2 A(x) A(y) 
+ 2 A(x) A(z) 
+ 2 A(y) A(z) 
\nonumber \\ &&
+(s + w - 2 x - 2 y - 2 z) A(w)
+(s + x - 2 w - 2 y - 2 z) A(x)
\nonumber \\ &&
+(s + y - 2 w - 2 x - 2 z) A(y)
+(s + z - 2 w - 2 x - 2 y) A(z)
\nonumber \\ &&
+ \frac{s^2}{6} - \frac{3}{4} s (w + x + y + z) - \frac{9}{4}(w^2 + x^2 + y^2 + z^2)
\nonumber \\ &&
+ 2 (w x + w y + w z + x y + x z + y z)
,
\\
Q^2 \frac{\partial}{\partial Q^2} I_{5a}(v,w,x,y,z) &=& 
S(v,w,x) + S(v,y,z) + A(w) + A(x) + A(y) + A(z) 
\nonumber \\ &&
+v - 2w -2x-2y-2z + s/4,
\\
Q^2 \frac{\partial}{\partial Q^2} I_{5b}(v,w,x,y,z) &=&
S(v,w,x) + I(v,y,z) +  A(w) + A(x) + A(y) + A(z) 
\nonumber \\ &&
+v - 2w -2x-2y-2z + 5 s/8,
\\
Q^2 \frac{\partial}{\partial Q^2} I_{5c}(v,w,x,y,z) &=&
\Bigl [ A(x) + A(y) + A(z) -x -  y - z + w/2\Bigr ] B(v,w)
+ A(x) 
\nonumber \\ &&
+ A(y) + A(z) - A(v)/2 - x - y - z + 9 (v+w)/8 - s/4.
\eeq
It is crucial that only the renormalized (non-boldfaced) master integrals appear in renormalized expressions for physical observables (for examples, see refs.~\cite{Ford:1992pn,Martin:2001vx,Martin:2005eg,Martin:2005ch,Martin:2014cxa,Martin:2015lxa,Martin:2015rea,Martin:2016xsp,Martin:2017lqn,Martin:2022qiv}), and therefore require numerical evaluation.

The results below involve polynomials that encode the threshold structure of the integrals, and which appear as denominators in derivatives of the master integrals. They are the triangle function,
\beq
\Delta (x,y,z) &=& (\sqrtx - \sqrty - \sqrtz) (\sqrtx + \sqrty - \sqrtz) (\sqrtx - \sqrty + \sqrtz) (\sqrtx + \sqrty + \sqrtz)
\\ &=& x^2 + y^2 + z^2 - 2 x y - 2 x z - 2 y z
\eeq
and the corresponding kinematic threshold function with four arguments,
\beq
\Psi (w,x,y,z) &=& 
(\sqrtw - \sqrtx - \sqrty - \sqrtz) (\sqrtw + \sqrtx - \sqrty - \sqrtz) 
(\sqrtw - \sqrtx + \sqrty - \sqrtz) 
\nonumber \\ && 
(\sqrtw + \sqrtx + \sqrty - \sqrtz) 
(\sqrtw - \sqrtx - \sqrty + \sqrtz)
(\sqrtw + \sqrtx - \sqrty + \sqrtz) 
\nonumber \\ && 
(\sqrtw - \sqrtx + \sqrty + \sqrtz) (\sqrtw + \sqrtx + \sqrty + \sqrtz)
\\
&=&
w^4 + x^4 + y^4 + z^4 
- 4 \bigl (
  w^3 x
+ w^3 y 
+ w^3 z 
+ w x^3 
+ w y^3 
+ w z^3 
+ x^3 y 
\nonumber \\ && 
+ x^3 z 
+ x y^3 
+ x z^3 
+ y^3 z 
+ y z^3 
\bigr ) 
+ 4 \bigl (
w^2 x y 
+ w^2 x z 
+ w^2 y z 
+ w x^2 y 
\nonumber \\ && 
+ w x^2 z 
+ w x y^2 
+ w x z^2 
+ w y^2 z 
+ w y z^2 
+ x^2 y z 
+ x y^2 z 
+ x y z^2 
\bigr )
\nonumber \\ && 
+ 6 \bigl (
  w^2 x^2 
+ w^2 y^2 
+ x^2 y^2 
+ w^2 z^2 
+ x^2 z^2 
+ y^2 z^2 
\bigr )
- 40 w x y z 
,
\eeq
and the threshold function with five arguments:
\beq
\Omega(s,w,x,y,z) &=& 
 (\sqrts - \sqrtw - \sqrtx - \sqrty - \sqrtz) (\sqrts + \sqrtw - \sqrtx - \sqrty - \sqrtz)
 \nonumber \\ && 
 (\sqrts - \sqrtw + \sqrtx - \sqrty - \sqrtz) (\sqrts + \sqrtw + \sqrtx - \sqrty - \sqrtz)
 \nonumber \\ && 
 (\sqrts - \sqrtw - \sqrtx + \sqrty - \sqrtz) (\sqrts + \sqrtw - \sqrtx + \sqrty - \sqrtz)
 \nonumber \\ && 
(\sqrts - \sqrtw + \sqrtx + \sqrty - \sqrtz) (\sqrts + \sqrtw + \sqrtx + \sqrty - \sqrtz)
\nonumber \\ && 
 (\sqrts - \sqrtw - \sqrtx - \sqrty + \sqrtz) (\sqrts + \sqrtw - \sqrtx - \sqrty + \sqrtz)
\nonumber \\ && 
 (\sqrts - \sqrtw + \sqrtx - \sqrty + \sqrtz) (\sqrts + \sqrtw + \sqrtx - \sqrty + \sqrtz)
\nonumber \\ && 
 (\sqrts - \sqrtw - \sqrtx + \sqrty + \sqrtz) (\sqrts + \sqrtw - \sqrtx + \sqrty + \sqrtz)
\nonumber \\ && 
 (\sqrts - \sqrtw + \sqrtx + \sqrty + \sqrtz) (\sqrts + \sqrtw + \sqrtx + \sqrty + \sqrtz)
 .
\eeq
Despite the appearances of square roots, this expands to a homogeneous polynomial of degree 8 in $s,w,x,y,z$, with 495 terms.

The numerators of expressions for derivatives of the master integrals contain many other complicated polynomials. The explicit form of these results are relegated to ancillary electronic files, suitable for use with computers. 

The derivatives of the 1-loop master integrals with respect to squared mass arguments are well-known:
\beq
{\bf A}(x') &=& (1 - \epsilon) {\bf A}(x)/x
,
\label{eq:dAdx}
\\
{\bf B}(x',y) &=& \Bigl [
(1 - 2 \epsilon) (x - y - s) {\bf B}(x,y)+ (1 - \epsilon) (x + y - s) {\bf A}(x)/x
\nonumber \\ && 
 +
2 (\epsilon - 1) {\bf A}(y) 
\Bigr ]/\Delta(s,x,y) ,
\label{eq:dBdx}
\eeq
For convenience, these and the more complicated known results for 
${\bf I}(x',y,z)$, 
${\bf S}(x',y,z)$, 
${\bf T}(x',y,z)$,
${\bf T}(x,y',z)$, 
${\bf F}(w',x,y,z)$, 
${\bf F}(w,x',y,z)$, 
${\bf G}(v',w,x,y,z)$,
and ${\bf G}(v,w',x,y,z)$ are provided in the ancillary file {\tt derivativesbold},
in computer-readable form.
Also given in that file are the derivatives with respect to $s$ of 
${\bf B}(x,y)$, ${\bf S}(x,y,z)$, ${\bf T}(x,y,z)$. All of the corresponding results for derivatives of
the renormalized integrals $A(x)$, $B(x,y)$, $I(x,y,z)$, $S(x,y,z)$, $T(x,y,z)$, $F(w,x,y,z)$, $G(v,w,x,y,z)$ with respect to the squared masses, $s$, and $Q^2$ are collected in the ancillary file {\tt derivatives}.  


In the following, master integrals are simply chosen as the ones that have unit numerators and the fewest possible number of propagators, with one exception in section \ref{sec:fourprop}.  That exception is made in order to eliminate an avoidable pseudo-threshold denominator factor in the differential equations. Other than that single exception, in the cases encountered in this paper, there are no arbitrary choices to be made, because of the generic masses.

\section{Expansions in small external momentum invariant\label{sec:expansions}}
\setcounter{equation}{0}
\setcounter{figure}{0}
\setcounter{table}{0} 
\setcounter{footnote}{1}

\begin{figure}[!t]
\begin{minipage}[]{0.4\linewidth}
\includegraphics[width=4.2cm,angle=0]{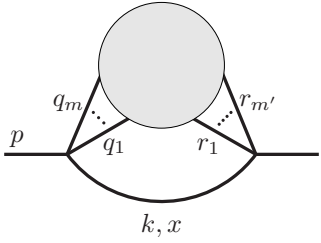}
\end{minipage}
\vspace{-0.4cm}
\begin{minipage}[]{0.55\linewidth}
\caption{\label{fig:easy}Diagram for a loop integral ${\bf f}(s;x,\ldots)$ with the property that the vertices where the two external legs are attached share an internal propagator with squared mass $x$ and momentum $k^\mu$. The external momentum invariant is $s = -p^2$. The small $s$ expansion for integrals of this type is given by eqs.~(\ref{eq:sexpansion})-(\ref{eq:definean}), in terms of derivatives  with respect to $x$ of the corresponding vacuum integral ${\bf f}(0;x,\ldots)$.}
\end{minipage}
\end{figure}

Consider the class of self-energy integrals in which at least one of the propagators connects the two vertices where the external legs are attached, as shown in Figure \ref{fig:easy}. 
Let the momentum and squared mass of this propagator be $k^\mu$ and $x$ respectively, and the external momentum is $p^\mu$ with invariant $s = -p^2$. The integral in question is denoted ${\bf f}(s;x,\ldots)$, with the dependence on the other internal squared masses indicated by the ellipses. The purpose of this section is to derive a simple formula for the small-$s$ expansion of ${\bf f}(s;x,\ldots)$, in terms of vacuum integrals, specifically the derivatives of ${\bf f}(0;x,\ldots)$ with respect to $x$, which are known for general masses up to 3-loop order \cite{Martin:2016bgz}.

To begin, let the other internal propagator momenta meeting at one of the external vertices be called $q_j^\mu$, with $j=1,\ldots,m$. Then the integral can be expressed as
\beq
{\bf f}(s;x,\ldots) &=& 
\int \frac{d^d\theta}{(2 \pi)^d} \int d^d k\> e^{i \theta \cdot (p - k - \sum_j q_j)}
\> 
G\>
\frac{1}{k^2 + x}
,
\eeq
where $G$ denotes the rest of the integral, and contains other propagators and momentum integrations, including integrations over the $q_j^\mu$, and can even have numerator factors, but has no direct dependence on $p^\mu$ or $k^\mu$. This allows us to write:
\beq
\frac{\partial}{\partial p_\mu}\frac{\partial}{\partial p^\mu}
{\bf f}(s;x,\ldots) &=& \int \frac{d^d\theta}{(2 \pi)^d} \int d^d k
\>
e^{i \theta \cdot (p - k - \sum_j q_j)}
\>
G 
\>
\frac{\partial}{\partial k_{\mu}}\frac{\partial}{\partial k^\mu} \frac{1}{k^2 + x} 
,
\eeq
which in turn can be expressed in terms of derivatives with respect to $x$. Doing this $n$ times gives
\beq
\left (\frac{\partial}{\partial p_\mu}\frac{\partial}{\partial p^\mu} \right )^n
{\bf f}(s;x,\ldots) 
&=&
\left ( - 4 x \frac{\partial^2}{\partial x^2} + 2 (d-4) \frac{\partial}{\partial x} \right )^n
{\bf f}(s;x,\ldots) 
.
\eeq
Now, using the identity
\beq
\left (\frac{\partial}{\partial p_\mu}\frac{\partial}{\partial p^\mu} \right )^n
(p^2)^n 
&=&
d(d+2)\ldots (d+2n-2)\, 2^n \, n! ,
\eeq
which can be verified by induction, I obtain the following simple power series in $s = -p^2$:
\beq
{\bf f}(s;x,\ldots) &=&
\sum_{n=0}^\infty
s^n a_n \hspace{1pt} {\cal D}_x^n \hspace{1.5pt} {\bf f}(0;x,\ldots),
\label{eq:sexpansion}
\eeq
where I have defined a differential operator:
\beq
{\cal D}_x &=& x \frac{\partial^2}{\partial x^2} + \epsilon \frac{\partial}{\partial x},
\label{eq:defineDx}
\eeq
and the coefficients appearing in the expansion are:
\beq
a_n &=& \frac{1}{n!} \frac{\Gamma(2 - \epsilon)}{\Gamma(n+2-\epsilon)}.
\label{eq:definean}
\eeq
Because derivatives of vacuum integrals ${\bf f}(0; x,\ldots)$ 
with respect to squared mass arguments are relatively easy
to find (see ref.~\cite{Martin:2016bgz} for the general case through 3-loop order), eqs.~(\ref{eq:sexpansion})-(\ref{eq:definean}) allows fast and straightforward evaluation of the small-$s$ expansion of all self-energy integrals of this class. This is the key result used to obtain the identities below.

In some cases, more than one of the internal masses can play the role of $x$ in the preceding discussion. Suppose that $x$ and $y$ are squared masses appearing in distinct single propagators that both directly connect the two external vertices. Then, because it
does not matter whether one uses ${\cal D}_x$ or ${\cal D}_y$ in
the expansion, one obtains the simple but non-trivial identity
\beq
{\cal D}_x\hspace{1pt} {\bf f}(s; x, y, \ldots) &=& {\cal D}_y \hspace{1pt}{\bf f}(s; x, y, \ldots)
.
\eeq 
For example, at one-loop order, one finds that the self-energy master integral obeys
\beq
{\cal D}_x \hspace{1pt}{\bf B}(x,y)
&=&
{\cal D}_y \hspace{1pt}{\bf B}(x,y),
\eeq
which can be checked using eqs.~(\ref{eq:dAdx}) and (\ref{eq:dBdx}).
Similarly, for the two-loop sunset integral,
\beq
{\cal D}_x \hspace{1pt}{\bf S}(x,y,z) \,=\, 
{\cal D}_y\hspace{1pt} {\bf S}(x,y,z) \,=\, 
{\cal D}_z \hspace{1pt}{\bf S}(x,y,z).
\eeq
This identity was noted in eq.~(3.7) of ref.~\cite{Martin:2003qz}, where it was expressed in terms of the renormalized version $S(x,y,z)$. Until now, the author had been somewhat perplexed by the existence of this identity, since it is not immediately obvious from the definition of the sunset integral or its symmetries. 

Similarly, for the 3-loop self-energy integrals considered in this paper, the above argument informs us that
\beq
{\cal D}_w \hspace{1pt}{\bf I}_4(w,x,y,z) \,=\, 
{\cal D}_x \hspace{1pt}{\bf I}_4(w,x,y,z) \,=\,
{\cal D}_y \hspace{1pt}{\bf I}_4(w,x,y,z) \,=\,
{\cal D}_z \hspace{1pt}{\bf I}_4(w,x,y,z),
\label{eq:J4secretsymmetry}
\eeq
and
\beq
{\cal D}_w \hspace{1pt}{\bf I}_{5b}(v,w,x,y,z) \,=\,
{\cal D}_x \hspace{1pt}{\bf I}_{5b}(v,w,x,y,z) ,
\label{eq:DwI5beqDxI5b}
\eeq
identities whose existence would otherwise be mysterious, at least to this author. 

\section{Three-loop four-propagator self-energy integrals\label{sec:fourprop}}
\setcounter{equation}{0}
\setcounter{figure}{0}
\setcounter{table}{0} 
\setcounter{footnote}{1}

\subsection{Inference of four-propagator self-energy integral identities from small $s$ expansions\label{subsec:fourpropsmalls}}

Consider the integral ${\bf I}_4(w,x,y,z)$. The expansion of this function to arbitrary order in $s$ can be obtained from eq.~(\ref{eq:sexpansion}) using ${\bf E}(w,x,y,z)$ in the role of ${\bf f}(0; x,\ldots)$. The derivatives with respect to $x$ are obtained using first eq.~(\ref{eq:defineboldF}) above, and then iteratively using the results 
for the derivatives of ${\bf F}(x,w,y,z)$
given originally in the ancillary file {\tt derivatives.txt} included with ref.~\cite{Martin:2016bgz}, and also provided in the ancillary file {\tt derivativesbold} of the present paper. Computing 
${\cal D}_x^n {\bf E}(w,x,y,z)$ in this way, I obtained the 
expansion to order $s^{24}$ of ${\bf I}_4(w,x,y,z)$. This was then used to obtain the expansions
for its first, second, and third derivatives with respect to the squared masses $w,x,y,z$. 
Then, plugging these into trial identities of the form eq.~(\ref{eq:genericidentity}), the polynomials giving valid identities between these integrals were solved for and checked, by considering for each power of $s$ the coefficients of each of the eight linearly independent vacuum master integrals ${\bf F}(w,x,y,z)$,
${\bf F}(x,w,y,z)$, ${\bf F}(y,w,x,z)$, ${\bf F}(z,w,x,y)$,
${\bf A}(w) {\bf A}(x) {\bf A}(y)$,
${\bf A}(w) {\bf A}(x) {\bf A}(z)$,
${\bf A}(w) {\bf A}(y) {\bf A}(z)$,
and
${\bf A}(x) {\bf A}(y) {\bf A}(z)$, and demanding that they vanish.

The simplest non-trivial such result involves the integral defined as follows:
\beq
{\bf J}_4 (w,x,y,z) &=& {\cal D}_w {\bf I}_4 (w,x,y,z) .
\label{eq:defineJ4}
\eeq
I find that this obeys
\beq
(s-w-x-y-z) {\bf J}_4 (w,x,y,z) &=&
\Bigl \{
(3 - 4 \epsilon) (2 - 3 \epsilon) + (6 \epsilon - 4) 
\bigl [ 
w \partial_w + x \partial_x + y \partial_y + z \partial_z
\bigr ]
\nonumber \\ &&
\!\!\!\!\!\!\!\!\!\!\!\!\!\!\!\!
\!\!\!\!\!\!\!\!\!\!\!\!\!\!\!\!
\!\!\!\!\!\!\!\!\!\!\!\!\!\!\!\!
\!\!\!\!\!\!\!\!\!\!\!\!\!\!\!\!
\!\!\!\!\!\!\!\!\!\!\!\!\!\!\!\!
+ 
2 \bigl [
w x \partial_w\partial_x +
w y \partial_w\partial_y +
w z \partial_w\partial_z +
x y \partial_x\partial_y +
x z \partial_x\partial_z +
y z \partial_y\partial_z 
\bigr ] 
\Bigr \} {\bf I}_4 (w,x,y,z).
\label{eq:JJ4identity}
\eeq
This identity has the very special feature that only the polynomial multiplying 
${\bf J}_4(w,x,y,z)$ involves $s$ at all, and it is linear in $s$. The fact that 
${\bf J}_4(w,x,y,z)$ is invariant under interchange of any of its arguments $w,x,y,z$ is not manifest from its definition eq.~(\ref{eq:defineJ4}), but is clear from eq.~(\ref{eq:JJ4identity}), in agreement with the argument leading to eq.~(\ref{eq:J4secretsymmetry}). 

Equation (\ref{eq:JJ4identity}) allows us to eliminate one of the integrals involved in it from the list of candidate master integrals. It is convenient to keep ${\bf J}_4(w,x,y,z)$ as a master integral,
and eliminate ${\bf I}_4(w,x,y,z)$ instead, because this prevents the appearance of factors of
$s-w-x-y-z$ in denominators of expressions for derivatives of the master integrals.
(This choice is made mainly for the sake of keeping the expressions as simple as possible. It also makes
the numerical evaluation more efficient for $s$ equal to, or very close to, $w+x+y+z$, but this is not crucial to get the numerical evaluation to work, as will be discussed further in section \ref{sec:numerical}.)
Also, the integrals 
${\bf I}_4(w'',x,y,z)$,
${\bf I}_4(x'',w,y,z)$,
${\bf I}_4(y'',w,x,z)$,
and
${\bf I}_4(z'',w,x,y)$
are all easily eliminated, because they can be written in terms of 
${\bf I}_4(w',x,y,z)$,
${\bf I}_4(x',w,y,z)$,
${\bf I}_4(y',w,x,z)$,
${\bf I}_4(z',w,x,y)$, and
${\bf J}_4(w,x,y,z)$, using
eqs.~(\ref{eq:J4secretsymmetry}) and (\ref{eq:defineJ4}).
I thus find that a good set of 4-propagator master integrals for generic $w,x,y,z$ can be chosen to be:
\beq
&&
{\bf J}_4 (w,x,y,z),\>\>\> 
{\bf I}_4 (w',x,y,z),\>\>\>
{\bf I}_4 (x',w,y,z),\>\>\>
{\bf I}_4 (y',w,x,z),\>\>\>
{\bf I}_4 (z',w,x,y),\>\>\>
\nonumber \\ &&
{\bf I}_4 (w',x',y,z),\>\>\>
{\bf I}_4 (w',y',x,z),\>\>\>
{\bf I}_4 (w',z',x,y),\>\>\>
\nonumber \\ &&
{\bf I}_4 (x',y',w,z),\>\>\>
{\bf I}_4 (x',z',w,y),\>\>\>
{\bf I}_4 (y',z',w,x),
\label{eq:II4masters}
\eeq
and the descendants of these integrals obtained by removing one propagator:
\beq
{\bf A}(w) {\bf A}(x) {\bf A}(y),\>\>\>
{\bf A}(w) {\bf A}(x) {\bf A}(z),\>\>\>
{\bf A}(w) {\bf A}(y) {\bf A}(z),\>\>\>
{\bf A}(x) {\bf A}(y) {\bf A}(z).
\label{eq:II4descendants}
\eeq

The derivatives of the master integrals in eq.~(\ref{eq:II4masters}) with respect to the squared mass arguments can now be obtained using the same strategy for constructing and verifying identities, as outlined in the Introduction. In the following, 
$\Omega \equiv \Omega(s,w,x,y,z)$. I find:
\beq
\Omega\, {\bf J}_4 (w',x,y,z) &=& 
(1 - 2 \epsilon) P_7\, {\bf J}_4 (w,x,y,z) 
+ (1 - 2 \epsilon) (2 - 3 \epsilon) P_7 \, {\bf I}_4 (w',x,y,z)
\nonumber \\ &&
+ (1 - 2 \epsilon) (2 - 3 \epsilon) \left[ 
P_6 \, x {\bf I}_4 (x',w,y,z) 
+ \{x \leftrightarrow y\}
+ \{x \leftrightarrow z\}
\right ]
\nonumber \\ &&
+ (1 - 2 \epsilon) \left[ 
P_7 \, x {\bf I}_4 (w',x',y,z) 
+ \{x \leftrightarrow y\}
+ \{x \leftrightarrow z\}
\right ]
\nonumber \\ &&
+ (1 - 2 \epsilon) \left[ 
P_6 \, x y {\bf I}_4 (x',y',w,z) 
+ \{x \leftrightarrow z\}
+ \{y \leftrightarrow z\}
\right ]
\nonumber \\ &&
+ (1 - \epsilon)^3 \left[ 
P_6 {\bf A}(x) {\bf A}(y) 
+ \{x \leftrightarrow z\}
+ \{y \leftrightarrow z\}
\right ] {\bf A}(w)/w
\nonumber \\ &&
+ (1 - \epsilon)^3 P_5 {\bf A}(x) {\bf A}(y) {\bf A}(z)
,
\label{eq:JJ4wpxyz}
\eeq
where each instance of $P_n$ indicates schematically the presence of a homogeneous polynomial in $w,x,y,z$ and $s$, of degree $n$ in the latter. 
Each such appearance of $P_n$, even within the same equation, stands for a different such polynomial, with the actual results found in the ancillary files. (In most cases, $n$ is also the squared mass dimension of $P_n$, but in a few cases the coefficient of $s^n$ is linear in the internal squared masses $v,w,x,\ldots$, so that the squared mass dimension of $P_n$ is $n+1$.)
Note that the dependences on $\epsilon$ have been factored out explicitly. Similarly, I find the schematic forms: 
\beq
\Omega\, {\bf I}_4 (w',x',y',z) &=& 
(1 - 2 \epsilon) P_6\,{\bf J}_4 (w,x,y,z) +
(1 - 2 \epsilon) (2 - 3 \epsilon) P_5 \, z {\bf I}_4 (z',w,x,y)
\nonumber \\ &&
+(1 - 2 \epsilon) (2 - 3 \epsilon) \left[ 
P_6 \, {\bf I}_4 (w',x,y,z) 
+ \{w \leftrightarrow x\}
+ \{w \leftrightarrow y\}
\right ]
\nonumber \\ &&
+ (1 - 2 \epsilon) \left[ 
P_7 \, {\bf I}_4 (w',x',y,z) 
+ \{w \leftrightarrow y\}
+ \{x \leftrightarrow y\}
\right ]
\nonumber \\ &&
+ (1 - 2 \epsilon) \left[ 
P_6 \, z {\bf I}_4 (w',z',x,y) 
+ \{w \leftrightarrow x\}
+ \{w \leftrightarrow y\}
\right ]
\nonumber \\ &&
+ (1 - \epsilon)^3 \left[ 
P_6\, {\bf A}(w) {\bf A}(x)/wx 
+ \{w \leftrightarrow y\}
+ \{x \leftrightarrow y\}
\right ] {\bf A}(z)
\nonumber \\ &&
+ (1 - \epsilon)^3 P_7\, {\bf A}(w) {\bf A}(x) {\bf A}(y)/wxy
,
\\
\Omega\, w {\bf I}_4 (w'',x',y,z)
&=& 
(1 - 2 \epsilon) P_7\,{\bf J}_4 (w,x,y,z)
+ (1 - 2 \epsilon) (2 - 3 \epsilon) P_6\, w {\bf I}_4 (w',x,y,z)
\nonumber \\ &&
+ (1 - 2 \epsilon) (2 - 3 \epsilon) P_7\, {\bf I}_4 (x',w,y,z)
\nonumber \\ &&
+(1 - 2 \epsilon) (2 - 3 \epsilon) \left[ 
P_6 \, y {\bf I}_4 (y',w,x,z) + \{ y \leftrightarrow z \} \right ]
\nonumber \\ &&
+ \left [(1 - 2 \epsilon) P_7 w   - \epsilon \Omega \right ] {\bf I}_4 (w',x',y,z)
+ (1 - 2 \epsilon) P_6\, y z {\bf I}_4 (y',z',w,x)
\nonumber \\ &&
+ (1 - 2 \epsilon) \left[ 
P_6 \, w y {\bf I}_4 (w',y',x,z) + \{y \leftrightarrow z\} \right ]
\nonumber \\ &&
+ (1 - 2 \epsilon) \left[ 
P_7 \, y {\bf I}_4 (x',y',w,z) + \{y \leftrightarrow z\} \right ]
\nonumber \\ && 
+ (1 - \epsilon)^3 \left [P_6\, {\bf A}(y) + \{y \leftrightarrow z\} \right ]
{\bf A}(w) {\bf A}(x)/x
\nonumber \\ &&
+ (1 - \epsilon)^3 P_5\,  {\bf A}(w) {\bf A}(y) {\bf A}(z)
+ (1 - \epsilon)^3 P_6\,  {\bf A}(x) {\bf A}(y) {\bf A}(z)/x
.
\label{eq:II4wppxpyz}
\eeq
The full explicit forms for eqs.~(\ref{eq:JJ4wpxyz})-(\ref{eq:II4wppxpyz}) are given in the ancillary file {\tt derivativesbold}.
These equations, applied recursively, enable one to find all higher derivatives with respect to the squared masses of the master integrals listed in eq.~(\ref{eq:II4masters}).

The derivatives of the master integrals with respect to $s$ can also be obtained from the preceding, by making use of the dimensional analysis constraint
\beq
s \frac{\partial}{\partial s} +
w \frac{\partial}{\partial w} +
x \frac{\partial}{\partial x} +
y \frac{\partial}{\partial y} +
z \frac{\partial}{\partial z} - n_p 
&=& 0,
\eeq
where $n_p$ is the squared-mass dimension of the integral being acted on, excluding the
$\mu$ dependence. (For example, 
$n_p=2 - 3 \epsilon$ for ${\bf I}_4$, and $n_p = 1 - 3 \epsilon$ for ${\bf J}_4$.)
The results are of the forms:
\beq
s \partials {\bf I}_4 (w',x,y,z) &=& (1 - 2 \epsilon) {\bf I}_4 (w',x,y,z) -{\bf J}_4(w,x,y,z)
- x {\bf I}_4 (w',x',y,z)
\nonumber \\ &&
- y {\bf I}_4 (w',y',x,z)
- z {\bf I}_4 (w',z',x,y)
,
\eeq
and
\beq
\Omega \, s \partials {\bf I}_4 (w',x',y,z) &=& 
(1 - 2 \epsilon) P_6\, s {\bf J}_4(w,x,y,z)
+ (1 - 2 \epsilon) (2 - 3 \epsilon)
\left [ P_7\, {\bf I}_4 (w',x,y,z) + \{w \leftrightarrow x\} \right ]
\nonumber \\ &&
+ (1 - 2 \epsilon) (2 - 3 \epsilon)
\left [P_6\, y {\bf I}_4 (y',w,x,z) + \{y \leftrightarrow z\} \right ]
\nonumber \\ &&
+\bigl [ (1 - 2 \epsilon) P_6 - \epsilon \Omega \bigr ] {\bf I}_4 (w',x',y,z)
+ (1 - 2 \epsilon) P_6 \, y z {\bf I}_4 (y',z',w,x)
\nonumber \\ &&
+ (1 - 2 \epsilon) \left (\left [ P_7\, y {\bf I}_4 (w',y',x,z) 
+ \{w \leftrightarrow x\}
\right ] 
+ \{y \leftrightarrow z\} \right )
\nonumber \\ && 
+ (1 - \epsilon)^3 \left [P_7  {\bf A}(y) + \{y \leftrightarrow z\} \right ] {\bf A}(w) {\bf A}(x)/wx
\nonumber \\ && 
+ (1 - \epsilon)^3 \left [P_6 {\bf A}(w) /w + \{w \leftrightarrow x\} \right ] {\bf A}(y) {\bf A}(z)
,
\eeq 
and
\beq
\Omega\, s \partials {\bf J}_4 (w,x,y,z) &=& 
\left [(1 - 3 \epsilon) \Omega + (1 - 2 \epsilon) P_7 \right ] {\bf J}_4 (w,x,y,z) 
\nonumber \\ &&
+ 
(1 - 2 \epsilon) (2 - 3 \epsilon) \Bigl [P_7\, w {\bf I}_4 (w',x,y,z) 
+ \{w \leftrightarrow x\}
+ \{w \leftrightarrow y\}
+ \{w \leftrightarrow z\}
\Bigr ]
\nonumber \\ &&
+ (1 - 2 \epsilon) \Bigl [ P_7\, w  x {\bf I}_4 (w',x',y,z) 
+ \mbox{(5 permutations)} \Bigr ]
\nonumber \\ &&
+  (1 - \epsilon)^3 \Bigl [ P_6\, {\bf A}(w) {\bf A}(x) {\bf A}(y) 
+ \{w \leftrightarrow z\}
+ \{x \leftrightarrow z\}
+ \{y \leftrightarrow z\}
\Bigr ]
.
\eeq
Again, the full explicit formulas are given in the ancillary file {\tt derivativesbold}.

For practical applications and numerical evaluation, it is appropriate to express results in terms of the renormalized (non-boldfaced) integrals
\beq
&&
{ J}_4 (w,x,y,z),\>\>\> 
{ I}_4 (w',x,y,z),\>\>\>
{ I}_4 (x',w,y,z),\>\>\>
{ I}_4 (y',w,x,z),\>\>\>
{ I}_4 (z',w,x,y),\>\>\>
\nonumber \\ &&
{ I}_4 (w',x',y,z),\>\>\>
{ I}_4 (w',y',x,z),\>\>\>
{ I}_4 (w',z',x,y),\>\>\>
\nonumber \\ &&
{ I}_4 (x',y',w,z),\>\>\>
{ I}_4 (x',z',w,y),\>\>\>
{ I}_4 (y',z',w,x),
\label{eq:I4mastersrenorm}
\eeq
defined by (\ref{eq:defineIXrenorm})-(\ref{eq:I43div})
along with the 1-loop integrals $A(w)$, $A(x)$, $A(y)$, and $A(z)$ defined by eqs.~(\ref{eq:defineArenorm}). 
Here the counterparts of eqs.~(\ref{eq:defineJ4})  and (\ref{eq:JJ4identity}) are the definition
\beq
J_4 (w,x,y,z) &=& w I_4 (w'',x,y,z) + A(w)/4 - 13 w/12,
\eeq
and the identity:
\beq
(s-w-x-y-z) J_4 (w,x,y,z) &=&
\Bigl \{
6 
- 4 
\bigl [ 
w \partial_w + x \partial_x + y \partial_y + z \partial_z
\bigr ]
\nonumber \\ &&
\!\!\!\!\!\!\!\!\!\!\!\!\!\!\!\!
\!\!\!\!\!\!\!\!\!\!\!\!\!\!\!\!
\!\!\!\!\!\!\!\!\!\!\!\!\!\!\!\!
\!\!\!\!\!\!\!\!\!\!\!\!\!\!\!\!
\!\!\!\!\!\!\!\!\!\!\!\!\!\!\!\!
+ 
2 \bigl [
w x \partial_w\partial_x +
w y \partial_w\partial_y +
w z \partial_w\partial_z +
x y \partial_x\partial_y +
x z \partial_x\partial_z +
y z \partial_y\partial_z 
\bigr ] 
\Bigr \} I_4 (w,x,y,z)
\nonumber \\ &&
\!\!\!\!\!\!\!\!\!\!\!\!\!\!\!\!
\!\!\!\!\!\!\!\!\!\!\!\!\!\!\!\!
\!\!\!\!\!\!\!\!\!\!\!\!\!\!\!\!
\!\!\!\!\!\!\!\!\!\!\!\!\!\!\!\!
\!\!\!\!\!\!\!\!\!\!\!\!\!\!\!\!
-A(w) A(x)
-A(w) A(y)
-A(w) A(z)
-A(x) A(y)
-A(x) A(z)
-A(y) A(z)
\nonumber \\ &&
\!\!\!\!\!\!\!\!\!\!\!\!\!\!\!\!
\!\!\!\!\!\!\!\!\!\!\!\!\!\!\!\!
\!\!\!\!\!\!\!\!\!\!\!\!\!\!\!\!
\!\!\!\!\!\!\!\!\!\!\!\!\!\!\!\!
\!\!\!\!\!\!\!\!\!\!\!\!\!\!\!\!
+(2 x + 2 y + 2 z - 3 w/4 - 5s/4) A(w)
+(2 w + 2 y + 2 z - 3 x/4 - 5s/4) A(x)
\nonumber \\ &&
\!\!\!\!\!\!\!\!\!\!\!\!\!\!\!\!
\!\!\!\!\!\!\!\!\!\!\!\!\!\!\!\!
\!\!\!\!\!\!\!\!\!\!\!\!\!\!\!\!
\!\!\!\!\!\!\!\!\!\!\!\!\!\!\!\!
\!\!\!\!\!\!\!\!\!\!\!\!\!\!\!\!
+(2 w + 2 x + 2 z - 3 y/4 - 5s/4) A(y)
+(2 w + 2 x + 2 y - 3 z/4 - 5s/4) A(z)
\nonumber \\ &&
\!\!\!\!\!\!\!\!\!\!\!\!\!\!\!\!
\!\!\!\!\!\!\!\!\!\!\!\!\!\!\!\!
\!\!\!\!\!\!\!\!\!\!\!\!\!\!\!\!
\!\!\!\!\!\!\!\!\!\!\!\!\!\!\!\!
\!\!\!\!\!\!\!\!\!\!\!\!\!\!\!\!
+ \bigl [ -25 s^2 + 102 s (w + x+ y + z) + 195 (w^2 + x^2 + y^2 + z^2) 
\nonumber \\ &&
\!\!\!\!\!\!\!\!\!\!\!\!\!\!\!\!
\!\!\!\!\!\!\!\!\!\!\!\!\!\!\!\!
\!\!\!\!\!\!\!\!\!\!\!\!\!\!\!\!
\!\!\!\!\!\!\!\!\!\!\!\!\!\!\!\!
\!\!\!\!\!\!\!\!\!\!\!\!\!\!\!\!
- 216 (w x + w y + w z
+ x y + x z + y z) \bigr ]/72
,
\label{eq:J4identity}
\eeq
which shows that $I_4(w,x,y,z)$ can be eliminated in favor of $J_4(w,x,y,z)$, thus avoiding
the appearance of $s-w-x-y-z$ in denominators, and also shows the non-trivial property that $J_4(w,x,y,z)$ is invariant under interchange of any two of $w,x,y,z$.
 
It then follows from the results above that the squared-mass derivatives of the renormalized master integrals are schematically of the forms:
\beq
\Omega\, J_4 (w',x,y,z) &=& 
P_7\, J_4 (w,x,y,z) 
+ P_7 \, I_4 (w',x,y,z)
\nonumber \\ &&
+ \left[ 
P_6 \, x I_4 (x',w,y,z) 
+ \{x \leftrightarrow y\}
+ \{x \leftrightarrow z\}
\right ]
\nonumber \\ &&
+ \left[ 
P_7 \, x I_4 (w',x',y,z) 
+ \{x \leftrightarrow y\}
+ \{x \leftrightarrow z\}
\right ]
\nonumber \\ &&
+ \left[ 
P_6 \, x y I_4 (x',y',w,z) 
+ \{x \leftrightarrow z\}
+ \{y \leftrightarrow z\}
\right ]
\nonumber \\ &&
+ \left[ 
P_6 A(x) A(y) 
+ \{x \leftrightarrow z\}
+ \{y \leftrightarrow z\}
\right ] A(w)/w
+ P_5 A(x) A(y) A(z)
\nonumber \\ &&
+ \left [P_7 A(x) 
+ \{x \leftrightarrow y\}
+ \{x \leftrightarrow z\}
\right ] A(w)/w
\nonumber \\ &&
+ \left [P_6 A(x) A(y) 
+ \{x \leftrightarrow z\}
+ \{y \leftrightarrow z\}
\right ] 
\nonumber \\ &&
+ P_8 A(w)/w 
+ \left [P_7 A(x) 
+ \{x \leftrightarrow z\} + \{y \leftrightarrow z\} \right ]
+ P_8
,
\label{eq:J4wpxyz}
\\
\Omega\, I_4 (w',x',y',z) &=& 
P_6\, J_4 (w,x,y,z) +
P_5 \, z I_4 (z',w,x,y)
\nonumber \\ &&
+ \left[ 
P_6 \, I_4 (w',x,y,z) 
+ \{w \leftrightarrow x\}
+ \{w \leftrightarrow y\}
\right ]
\nonumber \\ &&
+ \left[ 
P_7 \, I_4 (w',x',y,z) 
+ \{w \leftrightarrow y\}
+ \{x \leftrightarrow y\}
\right ]
\nonumber \\ &&
+ \left[ 
P_6 \, z I_4 (w',z',x,y) 
+ \{w \leftrightarrow x\}
+ \{w \leftrightarrow y\}
\right ]
\nonumber \\ &&
+ \left[ P_6 A(w) A(x)/wx 
+ \{w \leftrightarrow y\}
+ \{x \leftrightarrow y\}
\right ] A(z)
+ P_7 A(w) A(x) A(y)/wxy
\nonumber \\ &&
+ \left[ P_7 A(w) A(x)/wx 
+ \{w \leftrightarrow y\}
+ \{x \leftrightarrow y\}
\right ] 
\nonumber \\ &&
+ \left[ P_6 A(w)/w 
+ \{w \leftrightarrow x\}
+ \{w \leftrightarrow y\}
\right ] A(z) 
\nonumber \\ &&
+ \left[ P_7 A(w)/w 
+ \{w \leftrightarrow x\}
+ \{w \leftrightarrow y\}
\right ] 
+ P_6 A(z)
+ P_7
,
\\
\Omega\, w I_4 (w'',x',y,z)
&=& 
P_7\, J_4 (w,x,y,z)
+ P_6\, w I_4 (w',x,y,z)
+ P_7\, I_4 (x',w,y,z)
\nonumber \\ &&
+ \left[ P_6 \, y I_4 (y',w,x,z) + \{ y \leftrightarrow z \} \right ]
+ P_7 w I_4 (w',x',y,z)
\nonumber \\ &&
+ P_6\, y z I_4 (y',z',w,x)
+ \left[ 
P_6 \, w y I_4 (w',y',x,z) + (y \leftrightarrow z) \right ]
\nonumber \\ &&
+ \left[ 
P_7 \, y I_4 (x',y',w,z) + (y \leftrightarrow z) \right ]
+ \left [P_6\, A(y) + \{y \leftrightarrow z\} \right ] A(w) A(x)/x
\nonumber \\ &&
+ P_5\,  A(w) A(y) A(z)
+ P_6\,  A(x) A(y) A(z)/x
+ P_7\, A(w) A(x)/x
\nonumber \\ &&
+ \left [P_6\, A(y) + \{y \leftrightarrow z\} \right ] A(w)
+ \left [P_7\, A(y) + \{y \leftrightarrow z\} \right ] A(x)/x
\nonumber \\ &&
+ P_6\, A(y) A(z)
+ P_7\, A(w) 
+ P_8\, A(x)/x 
+ \left [P_7\, A(y) + \{y \leftrightarrow z\} \right ]
+ P_8
,\phantom{xxxx}
\eeq
while the derivatives with respect to $s$ are:
\beq
s \partials I_4 (w',x,y,z) &=& I_4 (w',x,y,z) - J_4(w,x,y,z)
- x I_4 (w',x',y,z)
- y I_4 (w',y',x,z)
\nonumber \\ &&
- z I_4 (w',z',x,y)
- \left [ A(x) + A(y) + A(z) + x + y + z - \frac{3w}{4} - \frac{s}{2} \right ] A(w)/w
\nonumber \\ &&
+ \frac{2w}{3} - \frac{s}{8}
,
\label{eq:sddsJ4wxyz}
\\
\Omega\, s \partials J_4 (w,x,y,z) &=& 
P_8 J_4 (w,x,y,z) 
+ 
\Bigl [P_7\, w I_4 (w',x,y,z) 
+ \{w \leftrightarrow x\}
+ \{w \leftrightarrow y\}
+ \{w \leftrightarrow z\}
\Bigr ]
\nonumber \\ &&
+ \Bigl [ P_7\, w  x I_4 (w',x',y,z) 
+ \mbox{(5 permutations)} \Bigr ]
\nonumber \\ &&
+  \Bigl [ P_6\, A(w) A(x) A(y) 
+ \{w \leftrightarrow z\}
+ \{x \leftrightarrow z\}
+ \{y \leftrightarrow z\}
\Bigr ]
\nonumber \\ &&
+  \Bigl [ P_7\, A(w) A(x) 
+ \mbox{(5 permutations)} \Bigr ]
\nonumber \\ &&
+  \Bigl [ P_8\, A(w) 
+ \{w \leftrightarrow x\}
+ \{w \leftrightarrow y\}
+ \{w \leftrightarrow z\}
\Bigr ]
+ P_9
,
\\
\Omega \, s \partials I_4 (w',x',y, z) &=& 
P_6\, s J_4(w,x,y,z)
+ \left [ P_7\, I_4 (w',x,y,z) + \{w \leftrightarrow x\} \right ]
\nonumber \\ &&
+ \left [P_6\, y I_4 (y',w,x,z) + \{y \leftrightarrow z\} \right ]
+ P_7\, I_4 (w',x',y,z)
+  P_6\, y z I_4 (y',z',w,x)
\nonumber \\ &&
+ \left (\left [ P_7\, y I_4 (w',y',x,z) 
+ \{w \leftrightarrow x\}
\right ]
+ \{y \leftrightarrow z\}
\right )
\nonumber \\ && 
+ \left [P_7\,  A(y) + \{y \leftrightarrow z\} \right ] A(w) A(x)/wx
\nonumber \\ && 
+ \left [P_6\, A(w) /w + \{w \leftrightarrow x\} \right ] A(y) A(z)
+ P_8\, A(w) A(x)/wx
\nonumber \\ && 
+ \left (\left [P_6\, A(w) A(y)/w 
+ \{w \leftrightarrow x\}
\right ]
+ \{y \leftrightarrow z\}
\right )
+ P_5\, A(y) A(z)
\nonumber \\ && 
+ \left [P_8\, A(w) /w + \{w \leftrightarrow x\} \right ]
+ \left [P_7\, A(y) + \{y \leftrightarrow z\} \right ]
+ P_8
\label{eq:sddsI4wpxpyz}
.
\eeq 
The full explicit expressions for eqs.~(\ref{eq:J4wpxyz})-(\ref{eq:sddsI4wpxpyz}) are given in the ancillary file {\tt derivatives}. Note that, as promised in ref.~\cite{Martin:2021pnd}, contributions of positive powers of $\epsilon$ in the expansions of ${\bf A}(x)$ etc.~do not appear. A further consistency check is provided by comparing the special case $w=x=y=z$ to the results obtained in ref.~\cite{Martin:2021pnd}.

Obtaining the numerical results for the renormalized master integrals is now straightforward,
using exactly the same method used for two-loop self-energy integrals in ref.~\cite{Martin:2005qm}. The coupled first-order differential equations (\ref{eq:sddsJ4wxyz})-(\ref{eq:sddsI4wpxpyz}) can be solved numerically, by applying a Runge-Kutta or similar method to integrate with respect to $s$ in the upper half complex plane, starting from $s=0$ using the boundary conditions
\beq
J_4(w,x,y,z) \Bigl |_{s=0} &=& -w F(w',x,y,z) + A(w)/4 - 13 w/12,
\\
I_4(w',x,y,z) \Bigl |_{s=0} &=& -F(w,x,y,z)
,
\label{eq:I4wpxyzbc}
\\
I_4(w',x',y,z) \Bigl |_{s=0} &=& -F(w,x',y,z)
,
\label{eq:I4wpxpyzbc}
\eeq
and obvious permutations thereof. The numerical values of the right sides of these boundary conditions can be evaluated using the results for the derivatives of $F$ in the ancillary file {\tt derivatives} and the {\tt 3VIL} code \cite{Martin:2016bgz}. For reasons of numerical stability, it is often better to start at a value slightly displaced from $s=0$, which can be done using the series expansions
implied by eq.~(\ref{eq:sexpansion}).

\subsection{Alternative method: expansions in one large mass\label{subsec:fourpropalternate}}

As noted near the end of the Introduction, the method of inferring identities using the polynomial form of coefficients resulting from IBP relations, without actually using the IBP procedure, can instead be carried out using other expansions (rather than small $s$). In this subsection I will briefly remark on a method that allows one to discover the
identities for the ${\bf B}(x,y)$ system at 1-loop, the 2-loop ${\bf S}(x,y,z)$ and ${\bf T}(x,y,z)$ system, and the 3-loop 4-propagator case, yielding the same results as in the previous subsection.

The idea is to choose one of the squared masses $z$ on a propagator connecting both external vertices as large, and to expand simultaneously in $s$ and all other squared masses. The tools necessary to find expansions of this type 
for all $N$-loop integrals with $N+1$ propagators were worked out in ref.~\cite{Berends:1993ee}. Applying the methods of that reference, one finds the completely analytic expansions valid when $z$ is large compared to $s,w,x,y$:
\beq
{\bf A}(z) &=& z \left (\frac{4 \pi \mu^2}{z}\right )^\epsilon \Gamma(\epsilon - 1) 
\\
{\bf B}(y,z) &=&  
\left (\frac{4 \pi \mu^2}{z}\right )^\epsilon \Gamma(\epsilon - 1) \Gamma(2 - \epsilon)
\sum_{n=0}^\infty \sum_{k=0}^\infty
\frac{(n+k)!}{n!\> k!\> \Gamma(n+2 - \epsilon)}
\left (\frac{s}{z} \right )^n
\left (\frac{y}{z} \right )^k
\nonumber \\ &&
\biggl [
\left (\frac{y}{z} \right )^{1 - \epsilon} \frac{\Gamma(n+k+2-\epsilon)}{\Gamma(k+2-\epsilon)}
- \frac{\Gamma(n+k+\epsilon)}{\Gamma(k+\epsilon)}
\biggr ]
\label{eq:BerendsB}
\\
{\bf S}(x,y,z) &=& 
z \left (\frac{4 \pi \mu^2}{z}\right )^{2 \epsilon} 
\left [\Gamma(\epsilon - 1) \Gamma(2 - \epsilon)\right ]^2 
\sum_{n=0}^\infty \sum_{k=0}^\infty \sum_{j=0}^\infty
\frac{\left ({s}/{z}\right)^n
      \left ({x}/{z}\right)^k
      \left ({y}/{z}\right)^j}{n!\>k!\>j!\>\Gamma(n+2 - \epsilon)}
\nonumber \\ &&      
\biggl [
\frac{\Gamma(j+k+n+\epsilon) \Gamma(j+k+n-1 + 2 \epsilon)}{\Gamma(k+\epsilon)\Gamma(j+\epsilon)}
\nonumber \\ &&      
- \left (\frac{y}{z} \right )^{1 - \epsilon} 
  \frac{\Gamma(j+k+n+1) \Gamma(j+k+n+ \epsilon)}{\Gamma(k+\epsilon)\Gamma(j+2-\epsilon)}     
\nonumber \\ &&
- \left (\frac{x}{z} \right )^{1 - \epsilon} 
  \frac{\Gamma(j+k+n+1) \Gamma(j+k+n+ \epsilon)}{\Gamma(k+2-\epsilon)\Gamma(j+\epsilon)}  
      \biggr ]
\nonumber \\ &&      
+ \left (\frac{x}{z} \right )^{1 - \epsilon} \left (\frac{y}{z} \right )^{1 - \epsilon}
  \frac{\Gamma(j+k+n+1) \Gamma(j+k+n+2- \epsilon)}{\Gamma(k+2-\epsilon)\Gamma(j+2-\epsilon)}  
      \biggr ]
\label{eq:BerendsS}
\\
{\bf I}_4(w,x,y,z) &=& 
-z^2 \left (\frac{4 \pi \mu^2}{z}\right )^{3 \epsilon} 
\left [\Gamma(\epsilon - 1) \Gamma(2 - \epsilon)\right ]^3
\sum_{n=0}^\infty \sum_{k=0}^\infty \sum_{j=0}^\infty \sum_{l=0}^\infty
\frac{\left ({s}/{z}\right)^n
      \left ({w}/{z}\right)^k
      \left ({x}/{z}\right)^j
      \left ({y}/{z}\right)^l}{n!\>k!\>j!\>l!\>\Gamma(n+2- \epsilon)}
\nonumber \\ &&      
\biggl [
\frac{\Gamma(j+k+l+n-2 + 3\epsilon) \Gamma(j+k+l+n-1 + 2 \epsilon)}{\Gamma(k+\epsilon)\Gamma(j+\epsilon)\Gamma(l+\epsilon)}
\nonumber \\ &&     
-
\left (\frac{y}{z} \right )^{1 - \epsilon}
\frac{\Gamma(j+k+l+n-1 + 2\epsilon) \Gamma(j+k+l+n+ \epsilon)}{\Gamma(k+\epsilon)\Gamma(j+\epsilon)\Gamma(l+2-\epsilon)}
\nonumber \\ && 
-
\left (\frac{x}{z} \right )^{1 - \epsilon}
\frac{\Gamma(j+k+l+n-1 + 2\epsilon) \Gamma(j+k+l+n+ \epsilon)}{\Gamma(k+\epsilon)\Gamma(j+2-\epsilon)\Gamma(l+\epsilon)}
\nonumber \\ &&   
 -
\left (\frac{w}{z} \right )^{1 - \epsilon}
\frac{\Gamma(j+k+l+n-1 + 2\epsilon) \Gamma(j+k+l+n+ \epsilon)}{\Gamma(k+2-\epsilon)\Gamma(j+\epsilon)\Gamma(l+\epsilon)}
\nonumber \\ &&
+ 
\left (\frac{y}{z} \right )^{1 - \epsilon} \left (\frac{x}{z} \right )^{1 - \epsilon}
\frac{\Gamma(j+k+l+n + \epsilon) \Gamma(j+k+l+n+ 1)}{\Gamma(k+\epsilon)\Gamma(j+2-\epsilon)\Gamma(l+2-\epsilon)}
\nonumber \\ &&
+ 
\left (\frac{y}{z} \right )^{1 - \epsilon} \left (\frac{w}{z} \right )^{1 - \epsilon}
\frac{\Gamma(j+k+l+n + \epsilon) \Gamma(j+k+l+n+ 1)}{\Gamma(k+2-\epsilon)\Gamma(j+\epsilon)\Gamma(l+2-\epsilon)}
\nonumber \\ &&
+ 
\left (\frac{x}{z} \right )^{1 - \epsilon} \left (\frac{w}{z} \right )^{1 - \epsilon}
\frac{\Gamma(j+k+l+n + \epsilon) \Gamma(j+k+l+n+ 1)}{\Gamma(k+2-\epsilon)\Gamma(j+2-\epsilon)\Gamma(l+\epsilon)}
\nonumber \\ &&
- 
\left (\frac{y}{z} \right )^{1 - \epsilon}
\left (\frac{x}{z} \right )^{1 - \epsilon} \left (\frac{w}{z} \right )^{1 - \epsilon}
\frac{\Gamma(j+k+l+n + 1) \Gamma(j+k+l+n+ 2-\epsilon)}{\Gamma(k+2-\epsilon)\Gamma(j+2-\epsilon)\Gamma(l+2-\epsilon)}
\biggr ]  
. 
\label{eq:BerendsI4}          
\eeq
For the three-loop 4-propagator case, arbitrary derivatives of 
${\bf I}_4(w,x,y,z)$ are immediately obtained from eq.~(\ref{eq:BerendsI4}). Then, consider a trial identity of the form of eq.~(\ref{eq:genericidentity}), with polynomials 
$C_k$ that are linear combinations of 
$s^{p_s} w^{p_w} x^{p_x} y^{p_y} z^{p_z}$, subject to the constraints that $p_s, p_w, p_x, p_y,$ and $p_z$ are all non-negative integers, with $p_s+ p_w+ p_x+ p_y+ p_z = n_k$. One can now
consider in turn each coefficient of a fixed power of $s, w, x, y$ in the identity, and require it to vanish, solving for one of the polynomial coefficients each time. Note that the power of $s$ in the identity is always a non-negative integer, while each of the powers of $w, x, y$ can be either an integer or an integer minus $\epsilon$, giving eight linearly independent constraints for each set of integer powers of $s,w,x,y$. By using eq.~(\ref{eq:BerendsI4}) truncated at large $n,k,l,j$, I have used this method to check the three-loop ${\bf I}_4$ topology identities claimed in the previous subsection. The same method applied to eqs.~(\ref{eq:BerendsB}) and (\ref{eq:BerendsS}) can be used to check the previously known identities for the 1-loop and 2-loop topologies.

I emphasize again that the validity of the identities obtained by this method does not rely on the convergence of the expansions for physically relevant values of $s$ and the squared masses. 
Once an identity has been put into polynomial coefficient form by multiplying by common denominators, one can even set $z=0$ with impunity, despite the fact that the expansion used to obtain it relied on the large $z$ limit (in this subsection) or the small $s$ limit (in the previous subsection).

\section{Three-loop five-propagator self-energy integrals\label{sec:fiveprop}}
\setcounter{equation}{0}
\setcounter{figure}{0}
\setcounter{table}{0} 
\setcounter{footnote}{1}

\subsection{Topology $I_{5a}$\label{subsec:I5a}}

Consider the self-energy integrals given by the topology ${\bf I}_{5a}$ shown in Figure \ref{fig:masters}. The small-$s$ expansion of the integral ${\bf I}_{5a}(v,w,x,y,z)$
can in principle be obtained to arbitrary order using eqs.~(\ref{eq:sexpansion})-(\ref{eq:definean}), with $v$ playing the role of $x$, and ${\bf G}(v,w,x,y,z)$ playing the role of ${\bf f}(0;v,\ldots)$. This can then be used to obtain the small-$s$ expansions of the derivatives of ${\bf I}_{5a}(v,w,x,y,z)$ with respect to its squared masses, in terms of the 17 linearly independent master vacuum integrals
\beq
&&
{\bf G}(v,w,x,y,z),\>\>\>\>\>
{\bf F}(w,x,y,z),\>\>\>\>\>
{\bf F}(x,w,y,z),\>\>\>\>\>
{\bf F}(y,w,x,z),\>\>\>\>\>
{\bf F}(z,w,x,y),
\nonumber \\ &&
{\bf A}(w) {\bf I}(v,y,z),\>\>\>\>\>
{\bf A}(x) {\bf I}(v,y,z),\>\>\>\>\>
{\bf A}(y) {\bf I}(v,w,x),\>\>\>\>\>
{\bf A}(z) {\bf I}(v,w,x),
\nonumber \\ &&
{\bf A}(v) {\bf A}(w) {\bf A}(y),\>\>\>\>\>
{\bf A}(v) {\bf A}(w) {\bf A}(z),\>\>\>\>\>
{\bf A}(v) {\bf A}(x) {\bf A}(y),\>\>\>\>\>
{\bf A}(v) {\bf A}(x) {\bf A}(z),
\nonumber \\ &&
{\bf A}(w) {\bf A}(x) {\bf A}(y),\>\>\>\>\>
{\bf A}(w) {\bf A}(x) {\bf A}(z),\>\>\>\>\>
{\bf A}(w) {\bf A}(y) {\bf A}(z),\>\>\>\>\>
{\bf A}(x) {\bf A}(y) {\bf A}(z).
\label{eq:II5avacuummasters}
\eeq
The results below were found and checked by doing the expansion to order $s^{20}$, using different rational numerical values of $v,w,x,y,z$ repeatedly in order to keep the sizes of the expressions small, until no further information could be obtained.

Then, using the method for discovering identities discussed in the Introduction,
I checked that the 5-propagator master integrals for this topology are: 
\beq
&&
{\bf I}_{5a}(v,w,x,y,z),\>\>\>\>\>
{\bf I}_{5a}(v',w,x,y,z),\>\>\>\>\>
{\bf I}_{5a}(v,w',x,y,z),
\nonumber \\ &&
{\bf I}_{5a}(v,x',w,y,z),\>\>\>\>\>
{\bf I}_{5a}(v,y',z,w,x),\>\>\>\>\>
{\bf I}_{5a}(v,z',y,w,x),
\label{eq:II5amasters}
\eeq
and their descendants obtained by removing one of the propagators:
\beq
&&
{\bf F}(w,x,y,z),\>\>\>\>\>
{\bf F}(x,w,y,z),\>\>\>\>\>
{\bf F}(y,w,x,z),\>\>\>\>\>
{\bf F}(z,w,x,y),
\nonumber \\ &&
{\bf A}(y) {\bf S}(v,w,x),\>\>\>\>\>
{\bf A}(y) {\bf T}(v,w,x),\>\>\>\>\>
{\bf A}(y) {\bf T}(w,v,x),\>\>\>\>\>
{\bf A}(y) {\bf T}(x,v,w),
\nonumber \\ &&
{\bf A}(z) {\bf S}(v,w,x),\>\>\>\>\>
{\bf A}(z) {\bf T}(v,w,x),\>\>\>\>\>
{\bf A}(z) {\bf T}(w,v,x),\>\>\>\>\>
{\bf A}(z) {\bf T}(x,v,w),
\nonumber \\ &&
{\bf A}(w) {\bf S}(v,y,z),\>\>\>\>\>
{\bf A}(w) {\bf T}(v,y,z),\>\>\>\>\>
{\bf A}(w) {\bf T}(y,v,z),\>\>\>\>\>
{\bf A}(w) {\bf T}(z,v,y),
\nonumber \\ &&
{\bf A}(x) {\bf S}(v,y,z),\>\>\>\>\>
{\bf A}(x) {\bf T}(v,y,z),\>\>\>\>\>
{\bf A}(x) {\bf T}(y,v,z),\>\>\>\>\>
{\bf A}(x) {\bf T}(z,v,y)
,
\label{eq:II5adescendants1}
\eeq
and further vacuum integral descendants ${\bf A}(v) {\bf A}(w) {\bf A}(y)$ etc., obtained by removing another propagator. The derivatives of the master integrals in eqs.~(\ref{eq:II5adescendants1}) were all previously known, and are given for completeness in the ancillary file {\tt derivativesbold}.

I then used the same method described in the Introduction to obtain the identities for the
derivatives of the master integrals in eq.~(\ref{eq:II5amasters}), as linear combinations
of the integrals in eqs.~(\ref{eq:II5amasters}) and (\ref{eq:II5adescendants1}).
The results for 
\beq
&&
{\bf I}_{5a}(v'',w,x,y,z),
\quad 
{\bf I}_{5a}(v',w',x,y,z),
\quad
\label{eq:II5amasterderivs}
\\ &&
{\bf I}_{5a}(v,w'',x,y,z),
\quad
{\bf I}_{5a}(v,w',x',y,z),
\quad
{\bf I}_{5a}(v,w',x,y',z),
\eeq
and permutations dictated by symmetries,
have the property that the coefficients are rational functions of
$v,w,x,y,z,s$, and $\epsilon$, with denominators involving
$\Psi(s,v,w,x)$, $\Psi(s,v,y,z)$, and $s-v$, but no other polynomials in $s$.
The derivatives of the master integrals with respect to $s$ are then obtained using dimensional analysis:
\beq
s \frac{\partial}{\partial s} {\bf I}_{5a}(v,w,x,y,z)
&=&
(1 - 3 \epsilon) {\bf I}_{5a}(v,w,x,y,z)
- v {\bf I}_{5a}(v',w,x,y,z)
- w {\bf I}_{5a}(v,w',x,y,z)
\nonumber \\ &&
- x {\bf I}_{5a}(v,x',w,y,z)
- y {\bf I}_{5a}(v,y',z,w,x)
- z {\bf I}_{5a}(v,z',y,w,x)
\label{eq:sddsII5a}
\\
s \frac{\partial}{\partial s} {\bf I}_{5a}(v',w,x,y,z)
&=&
- 3 \epsilon {\bf I}_{5a}(v',w,x,y,z)
- v {\bf I}_{5a}(v'',w,x,y,z)
- w {\bf I}_{5a}(v',w',x,y,z)
\nonumber \\ &&
- x {\bf I}_{5a}(v',x',w,y,z)
- y {\bf I}_{5a}(v',y',z,w,x)
- z {\bf I}_{5a}(v',z',y,w,x)
\\
s \frac{\partial}{\partial s} {\bf I}_{5a}(v,w',x,y,z)
&=&
- 3 \epsilon {\bf I}_{5a}(v,w',x,y,z)
- v {\bf I}_{5a}(v',w',x,y,z)
- w {\bf I}_{5a}(v,w'',x,y,z)
\nonumber \\ &&
- x {\bf I}_{5a}(v,w',x',y,z)
- y {\bf I}_{5a}(v,w',x,y',z)
- z {\bf I}_{5a}(v,w',x,z',y)
.\phantom{xxxxx}
\label{eq:sddsII5aw}
\eeq
The results for eqs.~(\ref{eq:II5amasterderivs})-(\ref{eq:sddsII5aw}) are given explicitly in terms of the master integrals in the ancillary file {\tt derivativesbold}.

From the above results, it is straightforward to obtain the corresponding non-trivial derivatives of the renormalized master integrals:
\beq
&&
{I}_{5a}(v'',w,x,y,z),
\qquad 
{I}_{5a}(v',w',x,y,z),
\qquad
\\ &&
{I}_{5a}(v,w'',x,y,z),
\qquad
{I}_{5a}(v,w',x',y,z),
\qquad
{I}_{5a}(v,w',x,y',z),
\label{eq:I5amasterderivs}
\\
&&
s \frac{\partial}{\partial s} {I}_{5a}(v,w,x,y,z),\qquad
s \frac{\partial}{\partial s} {I}_{5a}(v',w,x,y,z),\qquad
s \frac{\partial}{\partial s} {I}_{5a}(v,w',x,y,z),
\label{eq:sddsI5arenorm}
 \\
&&
Q^2 \frac{\partial}{\partial Q^2} {I}_{5a}(v,w,x,y,z),\qquad
Q^2 \frac{\partial}{\partial Q^2} {I}_{5a}(v',w,x,y,z),\qquad
Q^2 \frac{\partial}{\partial Q^2} {I}_{5a}(v,w',x,y,z)
.
\phantom{xxxxxxx}
\eeq
They are given in the ancillary file {\tt derivatives}.

The numerical evaluation of the renormalized master integrals
\beq
{}&&
{ I}_{5a}(v,w,x,y,z),\>\>\>
{ I}_{5a}(v',w,x,y,z),\>\>\>
{ I}_{5a}(v,w',x,y,z),\>\>\>
\nonumber \\ &&
{ I}_{5a}(v,x',w,y,z),\>\>\>
{ I}_{5a}(v,y',z,w,x),\>\>\>
{ I}_{5a}(v,z',y,w,x),\>\>\>
\nonumber \\ &&
S(v,w,x),\>\>\>T(v,w,x),\>\>\>T(w,v,x),\>\>\>T(x,v,w),
\nonumber \\ &&
S(v,y,z),\>\>\>T(v,y,z),\>\>\>T(y,v,z),\>\>\>T(z,v,y),
\label{eq:I5amasters}
\eeq
can now be accomplished by solving the coupled first-order differential equations in $s$.
The numerical solution by Runge-Kutta or a similar method starts from the boundary conditions at $s=0$ (or small s) in terms of the renormalized versions of the vacuum
integrals in eq.~(\ref{eq:II5avacuummasters}), which can be obtained from the results for the derivatives of $I$, $F$, and $G$ in the ancillary file {\tt derivatives}, and then using the code {\tt 3VIL}.

Besides the polynomials in $s$, the denominators of the expressions for ${\bf I}_{5a}(v,w'',x,y,z)$ and ${\bf I}_{5a}(v,w',x',y,z)$ contain factors of $w-x$ and $\Psi(w,x,y,z)$, which can vanish when $w=x$ and when $y=z$. The expression for ${\bf I}_{5a}(v,w',x,y',z)$ also has a factor of $\Psi(w,x,y,z)$. The same holds for the derivatives of the corresponding renormalized master integrals in eq.~(\ref{eq:I5amasterderivs}). In the special cases $w=x$ and $y=z$, the identities can be obtained by taking the corresponding limits. 
More importantly from a practical point of view, it should be noted that the $s$ derivatives of the master integrals in eq.~(\ref{eq:sddsI5arenorm}) are completely free of denominators that vanish when $w=x$ and/or $y=z$,
so that there is no obstacle to evaluating the master integrals numerically even in those special cases. In particular, I have checked that in the special case $v=w=x=y=z$, all of the results described above agree with those found (using the traditional IBP method) in ref.~\cite{Martin:2021pnd}.

\subsection{Topology $I_{5b}$}

Next, consider the self-energy integrals given by the topology ${\bf I}_{5b}$ shown in Figure \ref{fig:masters}. The small-$s$ expansion of the integral ${\bf I}_{5b}(v,w,x,y,z)$
can in principle be obtained to arbitrary order using eqs.~(\ref{eq:sexpansion})-(\ref{eq:definean}), with ${\bf f}(0;x,\ldots) = {\bf G}(v,w,x,y,z)$. This can then be used to obtain the small-$s$ expansions of arbitrary derivatives of ${\bf I}_{5b}(v,w,x,y,z)$ with respect to its squared mass arguments, in terms of the same 17 linearly independent master vacuum integrals that appeared in eq.~(\ref{eq:II5avacuummasters}). In practice, I found the results below using expansions to order $s^{20}$, repeatedly choosing different rational values for $v,w,x,y,z$ to keep the expressions tractable, until no further information could be obtained.

Doing so, I checked that the master integrals are:
\beq
&&
{\bf I}_{5b}(v,w,x,y,z),\>\>\>
{\bf I}_{5b}(v',w,x,y,z),\>\>\>
{\bf I}_{5b}(v,w',x,y,z),\>\>\>
{\bf I}_{5b}(v,x',w,y,z)
,
\label{eq:I5bmasters}
\eeq
along with their descendants obtained by removing one propagator, including the master integrals associated with the subsidiary topology ${\bf I}_4(w,x,y,z)$ found in eq.~(\ref{eq:II4masters}), as well as
\beq
&&
{\bf A}(y) {\bf S}(v,w,x),\>\>\>
{\bf A}(y) {\bf T}(v,w,x),\>\>\>
{\bf A}(y) {\bf T}(w,v,x),\>\>\>
{\bf A}(y) {\bf T}(x,w,v),\>\>\>
\nonumber \\ &&
{\bf A}(z) {\bf S}(v,w,x),\>\>\>
{\bf A}(z) {\bf T}(v,w,x),\>\>\>
{\bf A}(z) {\bf T}(w,v,x),\>\>\>
{\bf A}(z) {\bf T}(x,w,v),\>\>\>
\nonumber \\ &&
{\bf A}(w) {\bf I}(v,y,z),\>\>\>
{\bf A}(x) {\bf I}(v,y,z),
\label{eq:I5bdescendants1}
\eeq
and further vacuum integral descendants ${\bf A}(v) {\bf A}(w) {\bf A}(y)$ etc., obtained
by removing another propagator.

I then used the method described in the Introduction to obtain the identities yielding
the derivatives of the master integrals in eq.~(\ref{eq:I5bmasters}):
\beq
&&
{\bf I}_{5b}(v,w,x,y',z),
\quad 
{\bf I}_{5b}(v'',w,x,y,z),
\quad
{\bf I}_{5b}(v',w',x,y,z),
\quad
\nonumber \\ &&
{\bf I}_{5b}(v,w',x',y,z),
\quad
{\bf I}_{5b}(v,w'',x,y,z),
\label{eq:I5bderivatives}
\eeq
and others related to them by symmetries, as linear combinations of the master integrals. 
The first of these identities is particularly simple, as there are only a few terms and all of the polynomials are actually independent of $s$:
\beq
\Delta(v,y,z) {\bf I}_{5b}(v,w,x,y',z) &=&
(1 - 2 \epsilon) (y - v - z) {\bf I}_{5b}(v,w,x,y,z) 
+ (v-y-z) {\bf I}_{4}(y',w,x,z)
\nonumber \\ &&
\hspace{-80pt}
+ 2 z {\bf I}_{4}(z',w,x,y)
+ (1 - \epsilon) {\bf S}(v,w,x) \left [ (y+z-v) {\bf A}(y)/y - 2 {\bf A}(z) \right ].
\eeq
From the results for eq.~(\ref{eq:I5bderivatives}), all higher derivatives [such as ${\bf I}_{5b}(v''',w,x,y,z)$ and ${\bf I}(v',w,x,y',z)$] can be obtained by iteration, and the identity given above as eq.~(\ref{eq:DwI5beqDxI5b}) can be verified.
Furthermore, the derivatives with respect to $s$ are obtained using
\beq
s \frac{\partial}{\partial s} {\bf I}_{5b}(v,w,x,y,z)
&=&
(1 - 3 \epsilon) {\bf I}_{5b}(v,w,x,y,z)
- v {\bf I}_{5b}(v',w,x,y,z)
- w {\bf I}_{5b}(v,w',x,y,z)
\nonumber \\ &&
- x {\bf I}_{5b}(v,x',w,y,z)
- y {\bf I}_{5b}(v,w,x,y',z)
- z {\bf I}_{5b}(v,w,x,z',y)
,
\\
s \frac{\partial}{\partial s} {\bf I}_{5b}(v',w,x,y,z)
&=&
- 3 \epsilon {\bf I}_{5b}(v',w,x,y,z)
- v {\bf I}_{5b}(v'',w,x,y,z)
- w {\bf I}_{5b}(v',w',x,y,z)
\nonumber \\ &&
- x {\bf I}_{5b}(v',x',w,y,z)
- y {\bf I}_{5b}(v',w,x,y',z)
- z {\bf I}_{5b}(v',w,x,z',y)
,
\\
s \frac{\partial}{\partial s} {\bf I}_{5b}(v,w',x,y,z)
&=&
- 3 \epsilon {\bf I}_{5b}(v,w',x,y,z)
- v {\bf I}_{5b}(v',w',x,y,z)
- w {\bf I}_{5b}(v,w'',x,y,z)
\nonumber \\ &&
- x {\bf I}_{5b}(v,w',x',y,z)
- y {\bf I}_{5b}(v,w',x,y',z)
- z {\bf I}_{5b}(v,w',x,z',y)
.
\phantom{xxxxx}
\label{eq:sdsI5bw}
\eeq
The explicit results for eqs.~(\ref{eq:I5bderivatives})-(\ref{eq:sdsI5bw}) are given in the ancillary file {\tt derivativesbold}. Each of these results is a linear combination of the
master integrals in eqs.~(\ref{eq:I5bmasters})-(\ref{eq:I5bdescendants1}), with coefficients that are rational functions of $s,v,w,x,y,z$ and $\epsilon$, with denominator polynomials $\Psi(s,w,x,v)$ and $\Delta(v,y,z)$.

For the renormalized master integrals, the above results can be used to obtain the non-trivial derivatives 
\beq
&&
I_{5b}(v,w,x,y',z),
\quad 
I_{5b}(v'',w,x,y,z),
\quad
I_{5b}(v',w',x,y,z),
\quad
\nonumber \\ &&
I_{5b}(v,w',x',y,z),
\quad
I_{5b}(v,w'',x,y,z),
\\
&&
s \frac{\partial}{\partial s} I_{5b}(v,w,x,y,z),\qquad
s \frac{\partial}{\partial s} I_{5b}(v',w,x,y,z),\qquad
s \frac{\partial}{\partial s} I_{5b}(v,w',x,y,z),
\label{eq:sdsI5b}
\\
&&
Q^2 \frac{\partial}{\partial Q^2} I_{5b}(v,w,x,y,z),\qquad
Q^2 \frac{\partial}{\partial Q^2} I_{5b}(v',w,x,y,z),\qquad
Q^2 \frac{\partial}{\partial Q^2} I_{5b}(v,w',x,y,z)
,
\phantom{xxxxxxx}
\eeq
and others related by symmetries. They are given explicitly in the ancillary file {\tt derivatives}. I checked that in the special case $v=w=x=y=z$, all of the results described above agree with those found using the traditional IBP method in ref.~\cite{Martin:2021pnd}. The first-order coupled linear differential equations (\ref{eq:sdsI5b}), together with the ones listed in eq.~(\ref{eq:I4mastersrenorm})
and the ones for $S(v,w,x)$, $T(v,w,x)$, $T(w,v,x)$, $T(x,w,v)$, all listed in the
same ancillary file {\tt derivatives}, can be numerically solved simultaneously
using Runge-Kutta, as discussed above. 

\subsection{Topology $I_{5c}$}

Finally, consider the self-energy integrals given by the topology ${\bf I}_{5c}$ depicted in Figure \ref{fig:masters}. The small-$s$ expansion of the integral ${\bf I}_{5c}(v,w,x,y,z)$
can in principle be obtained to arbitrary order using eqs.~(\ref{eq:sexpansion})-(\ref{eq:definean}), with $v$ playing the role of $x$, and 
\beq
{\bf f}(0;v,\ldots) &=& \frac{{\bf E}(v,x,y,z) - {\bf E}(w,x,y,z)}{w-v} .
\eeq 
This can then be used to obtain the small-$s$ expansions of derivatives of 
${\bf I}_{5c}(v,w,x,y,z)$ with respect to its squared mass arguments, in terms of the
15 linearly independent master vacuum integrals
\beq
&&
{\bf F}(w,x,y,z),\>\>\>\>
{\bf F}(x,w,y,z),\>\>\>\>
{\bf F}(y,w,x,z),\>\>\>\>
{\bf F}(z,w,x,y),\>\>\>\>
\nonumber \\ &&
{\bf F}(v,x,y,z),\>\>\>\>
{\bf F}(x,v,y,z),\>\>\>\>
{\bf F}(y,v,x,z),\>\>\>\>
{\bf F}(z,v,x,y),\>\>\>\>
\nonumber \\ &&
{\bf A}(w) {\bf A}(x) {\bf A}(y),\>\>\>\>
{\bf A}(w) {\bf A}(x) {\bf A}(z),\>\>\>\>
{\bf A}(w) {\bf A}(y) {\bf A}(z),\>\>\>\>
{\bf A}(x) {\bf A}(y) {\bf A}(z),\>\>\>\>
\nonumber \\ &&
{\bf A}(v) {\bf A}(x) {\bf A}(y),\>\>\>\>
{\bf A}(v) {\bf A}(x) {\bf A}(z),\>\>\>\>
{\bf A}(v) {\bf A}(y) {\bf A}(z)
.
\eeq
In practice, I obtained the results below using expansions to order $s^{20}$, repeatedly choosing different rational values for $v,w,x,y,z$ until no further information could be obtained.

Doing so, I found that the master integrals for this topology are:
\beq
&&
{\bf I}_{5c}(v,w,x,y,z),\>\>\>\>\>
{\bf I}_{5c}(v,w,x',y,z),\>\>\>\>\>
{\bf I}_{5c}(v,w,y',x,z),\>\>\>\>\>
{\bf I}_{5c}(v,w,z',x,y)
,
\label{eq:II5cmasters}
\eeq
together with the ones for ${\bf I}_4(v,x,y,z)$, obtained from section \ref{subsec:fourpropsmalls} with $w \rightarrow v$,
and the other master integrals for descendants obtained by removing one of the propagators
in other ways:
\beq
&&
{\bf F}(w,x,y,z),\>\>\>\>\>
{\bf F}(x,w,y,z),\>\>\>\>\>
{\bf F}(y,w,x,z),\>\>\>\>\>
{\bf F}(z,w,x,y),\>\>\>\>\>
\nonumber \\ &&
{\bf A}(x) {\bf A}(y) {\bf B}(v,w),\>\>\>\>\>
{\bf A}(x) {\bf A}(z) {\bf B}(v,w),\>\>\>\>\>
{\bf A}(y) {\bf A}(z) {\bf B}(v,w).\>\>\>\>\>
\label{eq:II5cdescendants}
\eeq
Then, I used the method outlined in the Introduction to obtain expressions for the derivatives of the master integrals,
\beq
{\bf I}_{5c}(v',w,x,y,z),\>\>\>\>\>
{\bf I}_{5c}(v,w',x,y,z),\>\>\>\>\>
{\bf I}_{5c}(v,w,x'',y,z),\>\>\>\>\>
{\bf I}_{5c}(v,w,x',y',z),
\label{eq:II5cderivs}
\eeq
as linear combinations of the master integrals.
The first two can be written in a remarkably compact form, in terms of
$\Delta(s,v,w)$ (denoted as $\Delta$ in the remainder of this subsection):
\beq
\Delta\, {\bf I}_{5c}(v',w,x,y,z) 
&=&
(1 - 2 \epsilon) (v - w - s) {\bf I}_{5c}(v,w,x,y,z)
+ (w - 3v - s) {\bf I}_{4}(v',x,y,z)
\nonumber \\ &&
- 2 x {\bf I}_{4}(x',v,y,z)
- 2 y {\bf I}_{4}(y',v,x,z)
- 2 z {\bf I}_{4}(z',v,x,y)
\nonumber \\ &&
+ 2 (3 - 4 \epsilon) {\bf I}_4(v,x,y,z)
+ 2 (\epsilon - 1) {\bf E}(w,x,y,z)
,
\\
\Delta\, w {\bf I}_{5c}(v,w',x,y,z) 
&=&
(1 - 2 \epsilon)(s^2 - 2 s v - 3 s w + v^2 - 3 v w + 2 w^2)
{\bf I}_{5c}(v,w,x,y,z)
\nonumber \\ && 
- \Delta\, \bigl [
  x {\bf I}_{5c}(v,w,x',y,z)
+ y {\bf I}_{5c}(v,w,y',x,z)
+ z {\bf I}_{5c}(v,w,z',x,y) \bigr ]
\nonumber \\ && 
+ (3 - 4 \epsilon) (s-v-w) {\bf I}_4(v,x,y,z)
+ 2 (v-s) v {\bf I}_4(v',x,y,z)
\nonumber \\ && 
+ (v + w - s) \Bigl [ 
x {\bf I}_4(x',v,y,z)
+ y {\bf I}_4(y',v,x,z)
+ z {\bf I}_4(z',v,x,y)
\nonumber \\ && 
+ (1 - \epsilon) {\bf E}(w,x,y,z)
\Bigr ]
.
\eeq
Here I have used eqs.~(\ref{eq:EEFFidentity}) and (\ref{eq:JJ4identity}) to make the formulas even more compact. For the remaining two quantities in eq.~(\ref{eq:II5cderivs}),
the coefficients of the master integrals are somewhat more complicated but do not depend on $s$ at all, and have denominator polynomials $\Psi(w,x,y,z)$. The results for derivatives indicated in eq.~(\ref{eq:II5cderivs}) are provided in the ancillary file {\tt derivativesbold}.

The results for the derivatives of the master integrals with respect to $s$ follow from
dimensional analysis, and are simple enough that they can be written on a few lines:
\beq
\Delta \,
s \frac{\partial}{\partial s} {\bf I}_{5c} (v,w,x,y,z)
&=&
\left \{ (1 - 2 \epsilon)[s (v+w) - (v-w)^2] -\epsilon \Delta \right \} {\bf I}_{5c} (v,w,x,y,z)
\nonumber \\ &&
\hspace{-90pt}
+ (3 s + v - w) v{\bf I}_4(v',x,y,z) 
+ (s + v - w) \Bigl [
(4 \epsilon - 3) {\bf I}_4(v,x,y,z) 
\nonumber \\ &&
\hspace{-90pt}
+ x {\bf I}_4(x',v,y,z) 
+ y {\bf I}_4(y',v,x,z) 
+ z {\bf I}_4(z',v,x,y)
+ (1 - \epsilon) {\bf E}(w,x,y,z)
\Bigr ]
,
\\
\Delta \,
s \frac{\partial}{\partial s} {\bf I}_{5c} (v,w,x',y,z)
&=&
\left \{ (1 - 2 \epsilon)[s (v+w) - (v-w)^2] -\epsilon \Delta \right \} {\bf I}_{5c}(v,w,x',y,z)
\nonumber \\ &&
\hspace{-90pt}
+ (3 s + v-w) v {\bf I}_4(v',x',y,z)
+ (s + v - w) \Bigl [
(3 \epsilon - 2) {\bf I}_4(x',v,y,z)
\nonumber \\ &&
\hspace{-90pt}
+ z {\bf I}_4(x',z',v,y) + y {\bf I}_4(x',y',v,z) 
+ {\bf J}_4(v,x,y,z)
+ (\epsilon - 1) {\bf F}(x,w,y,z) \Bigr ]
,
\eeq
and the obvious permutations obtained from the latter equation with $x \leftrightarrow y$
or $x \leftrightarrow z$.
For convenience, these are also included in the ancillary file {\tt derivativesbold} in computer readable form, but written directly in terms of the master integrals rather than ${\bf E}(w,x,y,z)$ and ${\bf I}_4(v,x,y,z)$.

The corresponding results for the renormalized master integrals,
\beq
&&
{I}_{5c}(v',w,x,y,z),\>\>\>
{I}_{5c}(v,w',x,y,z),\>\>\>
{I}_{5c}(v,w,x'',y,z),\>\>\>
{I}_{5c}(v,w,x',y',z),
\nonumber \\ &&
s\frac{\partial}{\partial s} {I}_{5c}(v,w,x,y,z),\>\>\>
s\frac{\partial}{\partial s} {I}_{5c}(v,w,x',y,z),\>\>\>
\nonumber \\ &&
Q^2\frac{\partial}{\partial Q^2} {I}_{5c}(v,w,x,y,z),\>\>\>
Q^2\frac{\partial}{\partial Q^2} {I}_{5c}(v,w,x',y,z),\>\>\>
\eeq
and others related to them by symmetries are given in the ancillary file {\tt derivatives}.
In particular, the derivatives of the master integrals with respect to $s$ are
simple enough to present explicitly here:
\beq
\Delta \,
s \frac{\partial}{\partial s} I_{5c} (v,w,x,y,z)
&=&
\bigl [s (v+w) - (v-w)^2 \bigr ] I_{5c} (v,w,x,y,z)
\nonumber \\ &&
\hspace{-90pt}
+ (3 s + v - w) v I_4(v',x,y,z) 
+ (s + v - w) \Bigl [
- 3 I_4(v,x,y,z) 
\nonumber \\ &&
\hspace{-90pt}
+ x I_4(x',v,y,z) 
+ y I_4(y',v,x,z) 
+ z I_4(z',v,x,y)
+ E(w,x,y,z)
\nonumber \\ &&
\hspace{-90pt}
+ A(v) A(x) + A(v) A(y) + A(v) A(z) + A(x) A(y) + A(x) A(z) + A(y) A(z)
\nonumber \\ &&
\hspace{-90pt}
+ (-x-y-z + v/2 + s/2) A(v) 
+ (-v-y-z + x/2 + s/4) A(x) 
\nonumber \\ &&
\hspace{-90pt}
+ (-v-x-z + y/2 + s/4) A(y) 
+ (-v-x-y + z/2 + s/4) A(z) 
\nonumber \\ &&
\hspace{-90pt}
+ v x + v y + v z + x y + x z + y z
-9 (v^2 + x^2 + y^2 + z^2)/8
\nonumber \\ &&
\hspace{-90pt}
- s ([23 v + 7 w]/24 + [x+y+z]/6)
+ 7 s^2/36
\Bigr ]
,
\\
\Delta \,
s \frac{\partial}{\partial s} I_{5c} (v,w,x',y,z)
&=&
\bigl [s (v+w) - (v-w)^2 \bigr ] I_{5c}(v,w,x',y,z)
\nonumber \\ &&
\hspace{-90pt}
+ (3 s + v - w) v I_4(v',x',y,z)
+ (s + v - w) \Bigl [
- 2 I_4(x',v,y,z)
\nonumber \\ &&
\hspace{-90pt}
+ z I_4(x',z',v,y) + y I_4(x',y',v,z) 
+ J_4(v,x,y,z)
- F(x,w,y,z) 
\nonumber \\ &&
\hspace{-90pt}
+ \bigl [A(v) + A(y) + A(z) -v-y-z + 3x/4 + s/4\bigr ] A(x)/x
-2 x/3 + s/12
\Bigr ]
.
\eeq
Note that these differential equations are free of denominator factors that could vanish identically when $w=x$ and $y=z$. These results, together with the derivatives of $B(v,w)$ and the master integrals for the 4-propagator topology with arguments $v,x,y,z$, as worked out in section \ref{subsec:fourpropsmalls}, can be used for numerical evaluation of the master integrals, as discussed above.
I have again checked that in the special case $v=w=x=y=z$, all of the results described above agree with those found using the traditional IBP method in ref.~\cite{Martin:2021pnd}.

\section{Numerical evaluation\label{sec:numerical}}
\setcounter{equation}{0}
\setcounter{figure}{0}
\setcounter{table}{0} 
\setcounter{footnote}{1}

\begin{figure}[!t]
\begin{minipage}[]{0.50\linewidth}
\includegraphics[width=7.5cm,angle=0]{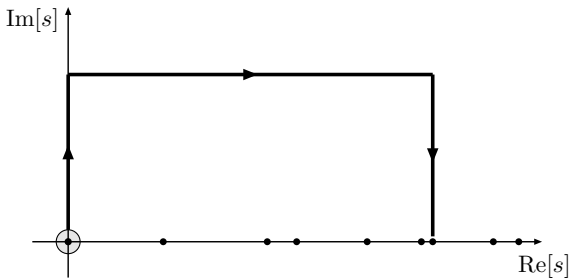}
\end{minipage}
\hspace{0.02\linewidth}
\begin{minipage}[]{0.45\linewidth}
\caption{\label{fig:contour}Contour for evaluation of self-energy master integrals by using their coupled first order differential equations in the external momentum invariant $s$. The initial boundary conditions are set at (or near) $s=0$ in terms of vacuum integrals
as in eqs.~(\ref{eq:Bvwinit})-(\ref{eq:I5c3init}), and then evolved by Runge-Kutta or similar methods along the path
in the upper-half complex $s$ plane, thus avoiding threshold and pseudo-threshold singularities indicated as dots 
on the real-$s$ axis.}
\end{minipage}
\end{figure}

As already mentioned above, one of the main reasons for obtaining the identities above is to enable the numerical computation of the master integrals. In general, one starts with the master integrals at (or near) $s=0$, using the values of vacuum integrals obtained by using, 
for example, the code {\tt 3VIL} \cite{Martin:2016bgz}. Then, the 
coupled first-order differential equations for master integrals $I_j(s)$ are of the form
\beq
\frac{d}{ds} I_j = \sum_{k} c_{jk}(s)\, I_k,
\eeq
which can be solved by Runge-Kutta or similar methods. The explicit forms of the differential equations are given in
the ancillary file {\tt derivatives}.
In order to get the branch cuts correct, a rectangular contour is chosen
in the upper-half complex $s$ plane to avoid threshold and pseudo-threshold singularities, as shown in Figure \ref{fig:contour}, as first suggested in ref.~\cite{Caffo:2002ch,Caffo:2002wm,Caffo:2003ma}. The height of the contour is arbitrary, and can be varied as a check on the numerical accuracy and stability. Because of the possibility that there may be a threshold or pseudo-threshold singularity at or near the desired final value of $s$, one should choose a Runge-Kutta algorithm that does not use calculation of the Runge-Kutta coefficients exactly at the final endpoint; a specific example of such an algorithm was provided in ref.~\cite{Martin:2005qm}, but there are many other such algorithms. To speed up the computation for a Runge-Kutta program with adaptive step size, and increase the accuracy for a fixed working precision, it is preferable to choose master integrals in such a way as to avoid singularities in the differential equations, to the extent possible. (We did this for the case of the topology $I_4(w,x,y,z)$, by avoiding the basis where a denominator $s-w-x-y-z$ would have appeared.) However, with arbitrary precision arithmetic and adaptive step-size control algorithms, any desired accuracy can in principle be obtained even if there are singular points on the real-$s$ line, at the cost of some computation time.

For the initial condition at $s=0$, the necessary boundary values for the master integrals treated in this paper are as follows:
\beq
B(v,w) |_{s=0} &=& [A(v) - A(w)]/(w-v),
\label{eq:Bvwinit}
\\
S(x,y,z) |_{s=0} &=& I(x,y,z),
\\
T(x,y,z) |_{s=0} &=& -I(x',y,z),
\\
I_4(w,x,y,z) |_{s=0} &=& E(w,x,y,z),
\\
I_4(w',x,y,z) |_{s=0} &=& -F(w,x,y,z),
\\
I_4(w',x',y,z) |_{s=0} &=& -F(w,x',y,z),
\\
J_4(w,x,y,z) |_{s=0} &=& -w F(w',x,y,z) + A(w)/4 - 13 w/12,
\\
I_{5a}(v,w,x,y,z) |_{s=0} &=& G(v,w,x,y,z),
\\
I_{5a}(v',w,x,y,z) |_{s=0} &=& G(v',w,x,y,z),
\\
I_{5a}(v,w',x,y,z) |_{s=0} &=& G(v,w',x,y,z),
\\
I_{5b}(v,w,x,y,z) |_{s=0} &=& G(v,w,x,y,z),
\\
I_{5b}(v',w,x,y,z) |_{s=0} &=& G(v',w,x,y,z),
\\
I_{5b}(v,w',x,y,z) |_{s=0} &=& G(v,w',x,y,z),
\\
I_{5c}(v,w,x,y,z) |_{s=0} &=& [E(v,x,y,z) - E(w,x,y,z)]/(w-v),
\label{eq:I5cinit}
\\
I_{5c}(v,w,x',y,z) |_{s=0} &=& [F(x,w,y,z) - F(x,v,y,z)]/(w-v),
\label{eq:I5c3init}
\eeq
The derivatives of the vacuum master integral on the right-hand sides of these equations can be obtained in terms of the vacuum master integrals, using the results presented in the same notation in the ancillary file {\tt derivatives.txt} of ref.~\cite{Martin:2016bgz}. For $v=w$, eqs.~(\ref{eq:Bvwinit}), (\ref{eq:I5cinit}), and (\ref{eq:I5c3init}) have singular denominators, but the limits are smooth:
\beq
B(w,w) |_{s=0} &=& -1 - A(w)/w,
\\
I_{5c}(w,w,x,y,z) |_{s=0} &=& F(w,x,y,z),
\\
I_{5c}(w,w,x',y,z) |_{s=0} &=& F(w,x',y,z).
\eeq
The non-generic case of masses $x,x,y,y$ for the 4-propagator vacuum integrals requires some care, as it
corresponds to the somewhat less trivial combined limit $w \rightarrow x$ and $z \rightarrow y$, for which I now present the results necessary for their evaluation. First, one has the identity
\beq
F(y,y,x,x) + F(x,x,y,y) &=&  
[A(x) + A(y) - 2 (x+y)] A(x) A(y)/xy + A(x)^2/x + A(y)^2/y 
\nonumber \\ &&
+ [2y/x -15/4] A(x) + [2 x/y - 15/4] A(y) + 14 (x+y)/3.
\label{eq:Fyyxxxxyy}
\eeq
Then one has the derivative formulas
\beq
4 x F(x',x,y,y) &=& -G(0,x,x,y,y) 
+ \frac{x+y}{x-y} F(x,x,y,y) 
+ \frac{2}{x (y-x)} A(x)^2 A(y)
\nonumber \\ && 
+ \frac{1}{x y} A(x) A(y)^2 
+ \frac{2}{x} A(x)^2 
+ \frac{x-3 y}{y (x-y)} A(y)^2
+ \frac{2(x+y)}{x (x-y)} A(x) A(y)
\nonumber \\ && 
+ \frac{3 x^2 + 3 x y - 8 y^2}{4 x (x-y)} A(x)
+ \frac{4 y}{x-y} A(y)
+ \frac{17 x^2 + x y + 10 y^2}{3 (y-x)}
,
\\
4 x F(x,x',y,y) &=& G(0,x,x,y,y) 
+ \frac{3x-y}{x-y} F(x,x,y,y) 
+ \frac{2}{x (y-x)} A(x)^2 A(y)
\nonumber \\ && 
- \frac{1}{x y} A(x) A(y)^2 
+ \frac{2}{x} A(x)^2 
+ \frac{x+y}{y (y-x)} A(y)^2
+ \frac{2(3x-y)}{x (x-y)} A(x) A(y)
\nonumber \\ && 
+ \frac{3 x^2 + 7 x y - 8 y^2}{4 x (y-x)} A(x)
+ \frac{4 y}{x-y} A(y)
+ \frac{4 x^2 + 10 x y + 14 y^2}{3 (y-x)}
,
\\
(x-y) F(x,y',x,y) &=&  -F(x,x,y,y)/2 + A(x)^2 A(y)/2 x y + A(y)^2/2 y
- A(x) A(y)/x
\nonumber \\ && 
+ (y/x - 7/8) A(x) - A(y) + 4x/3 + y
.
\label{eq:Fxypxy}
\eeq
The integrals $F(x,x,y,y)$ and $G(0,x,x,y,y)$ appearing in eqs.~(\ref{eq:Fyyxxxxyy})-(\ref{eq:Fxypxy}) are given in terms of polylogarithms in ref.~\cite{Martin:2016bgz}, and so can be very quickly evaluated to arbitrary accuracy.
The further limit $y \rightarrow x$ is also smooth: 
\beq
F(x,x,x,x) &=& A(x)^3/x^2 - A(x)^2/x - 7 A(x)/4 + 14x/3,
\\
F(x',x,x,x) &=& 2 A(x)^2/x^2 - 15 A(x)/4x + 35/12 - 7 \zeta_3,
\\
F(x,x',x,x) &=& A(x)^3/3 x^3 + 7 \zeta_3/3.
\eeq

In general, for faster performance, one can also use initial boundary conditions at a small non-zero value of $s$,
obtained by deriving the power series solution to the differential equation in $s$ using the results above for the $s^0$ terms. (Here it is important that the initial value of $s$ is not above, or close to, the lowest threshold of the integral. In particular, it is assumed that $s=0$ is not a threshold; otherwise terms involving $\ln(-s)$ would be necessary in the expansion.)

For the master integrals studied in this paper, I have used the out-of-the-box differential equation solver {\tt NDSolve} in Mathematica as a proof-of-principle for the numerical evaluation. (The same method was used for the 3-loop vacuum integrals that were used as the boundary conditions.)
This is not particulary fast, but allows for arbitrary numerical precision by a suitable choice of the {\tt WorkingPrecision} parameter. A few minutes total computing time is needed with a single 4.2 GHz processor to obtain 24 digits of precision for all of the 5-propagator topologies at a fixed $s$, with somewhat longer times needed when $s$ is at (or very close to) a threshold, and shorter times needed when $s$ is smaller. Note that the differential equations method computes simultaneously all of the relevant master integrals for a given topology and its sub-topologies. A much more efficient and optimized dedicated code is certainly possible, and may appear after the corresponding results for 6-, 7-, and 8-propagator master integrals become available.

\begin{figure}[!p]
\begin{center}
\includegraphics[width=0.38\linewidth,angle=0]{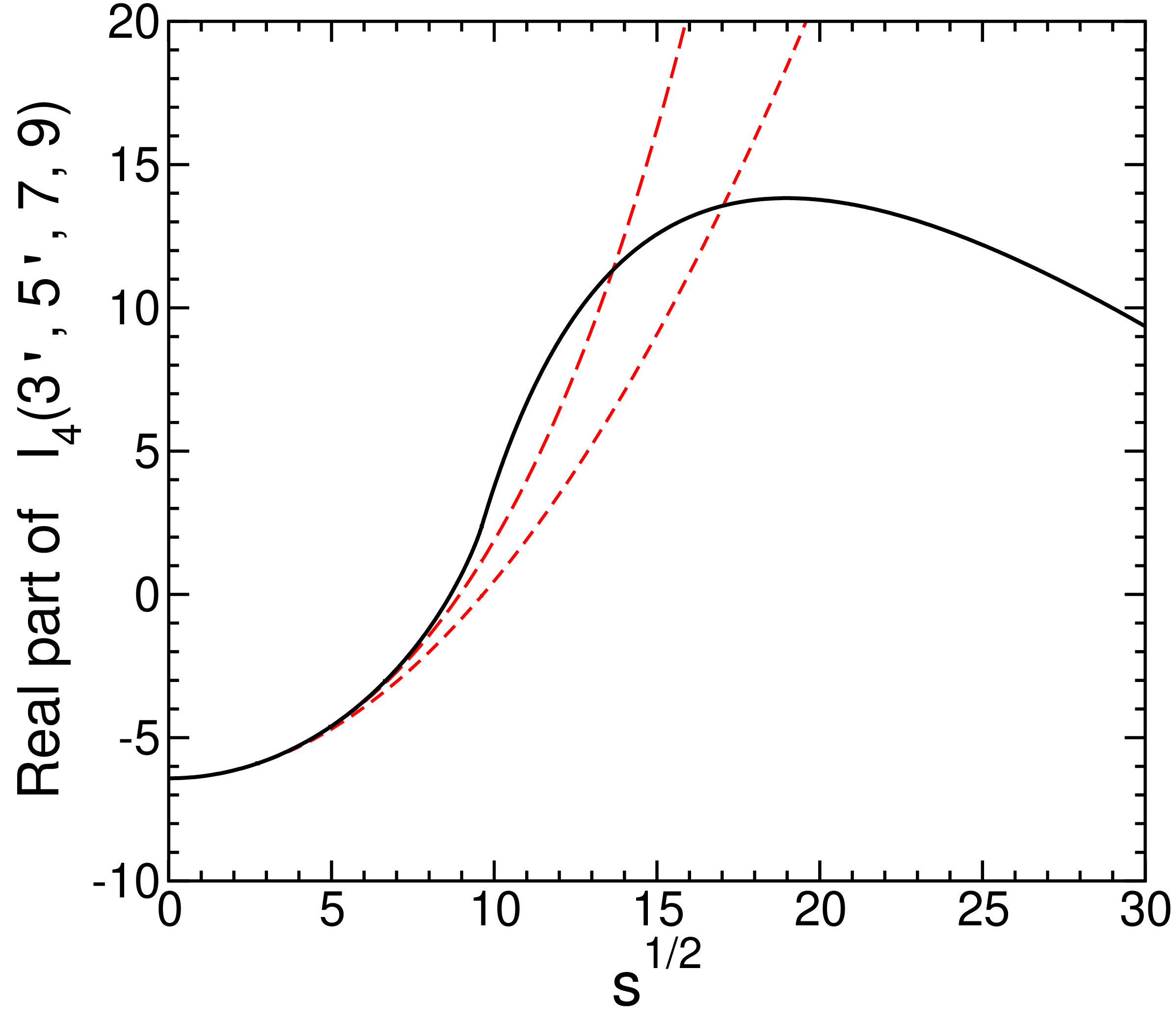}
\includegraphics[width=0.38\linewidth,angle=0]{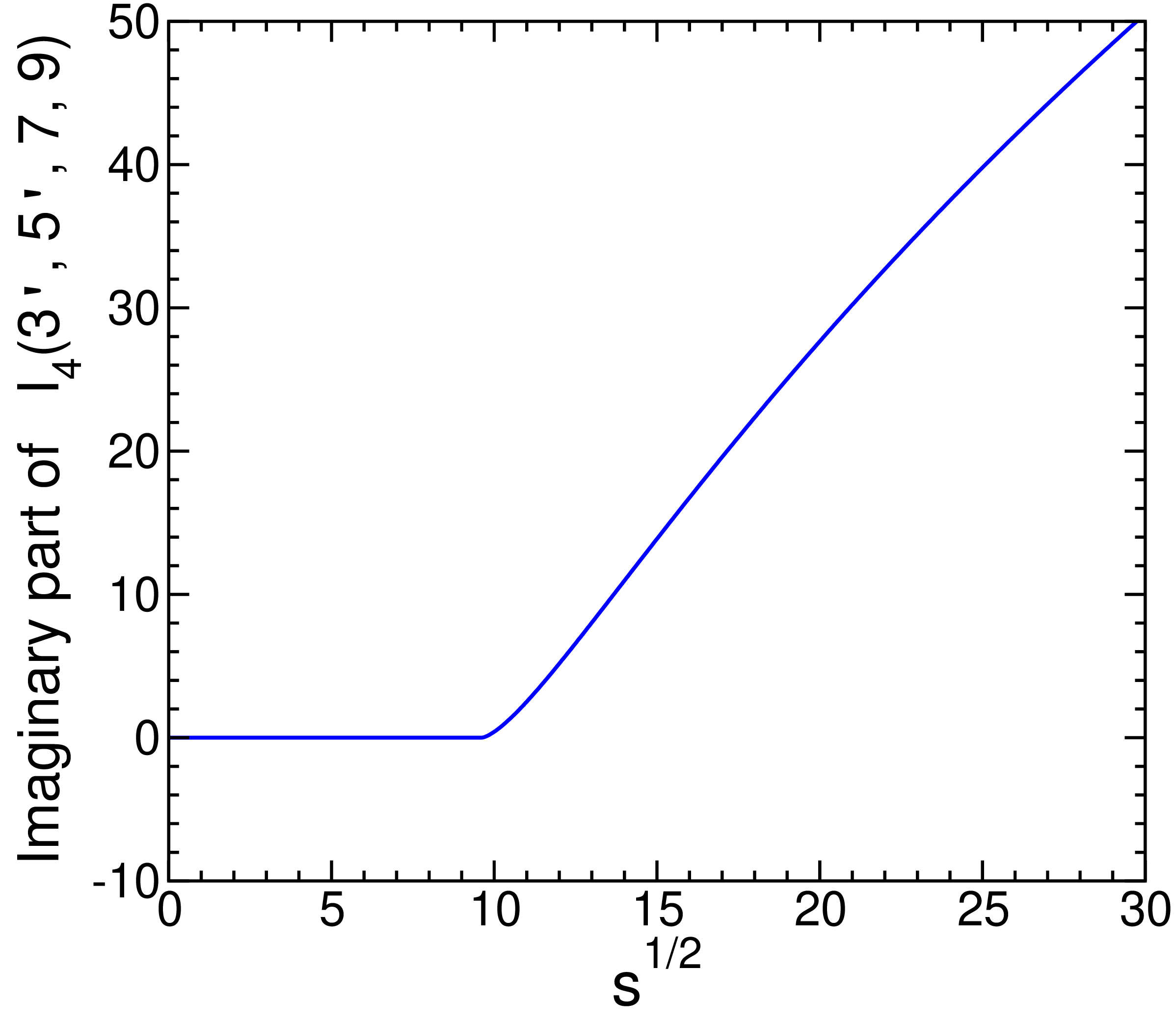}
\includegraphics[width=0.38\linewidth,angle=0]{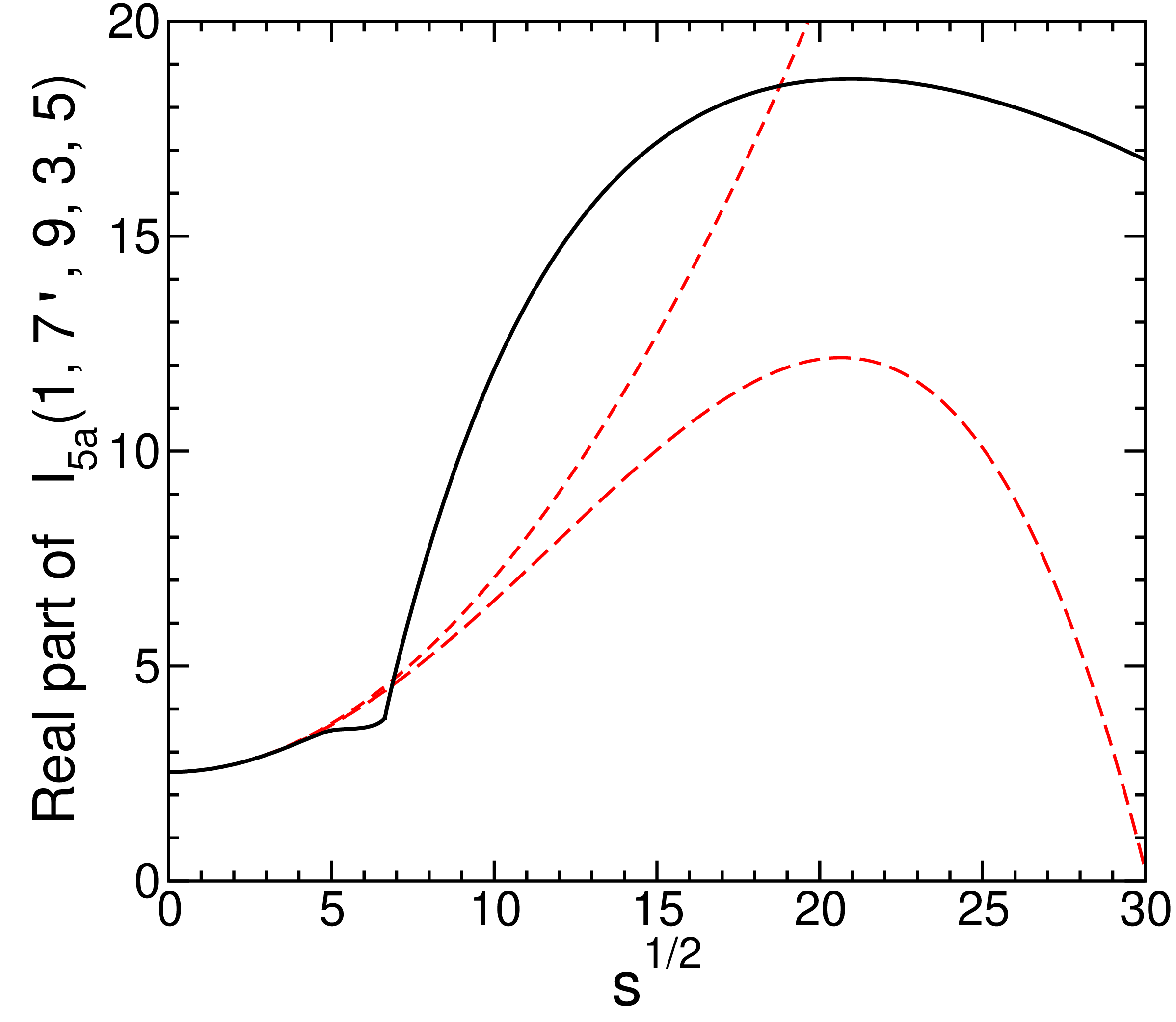}
\includegraphics[width=0.38\linewidth,angle=0]{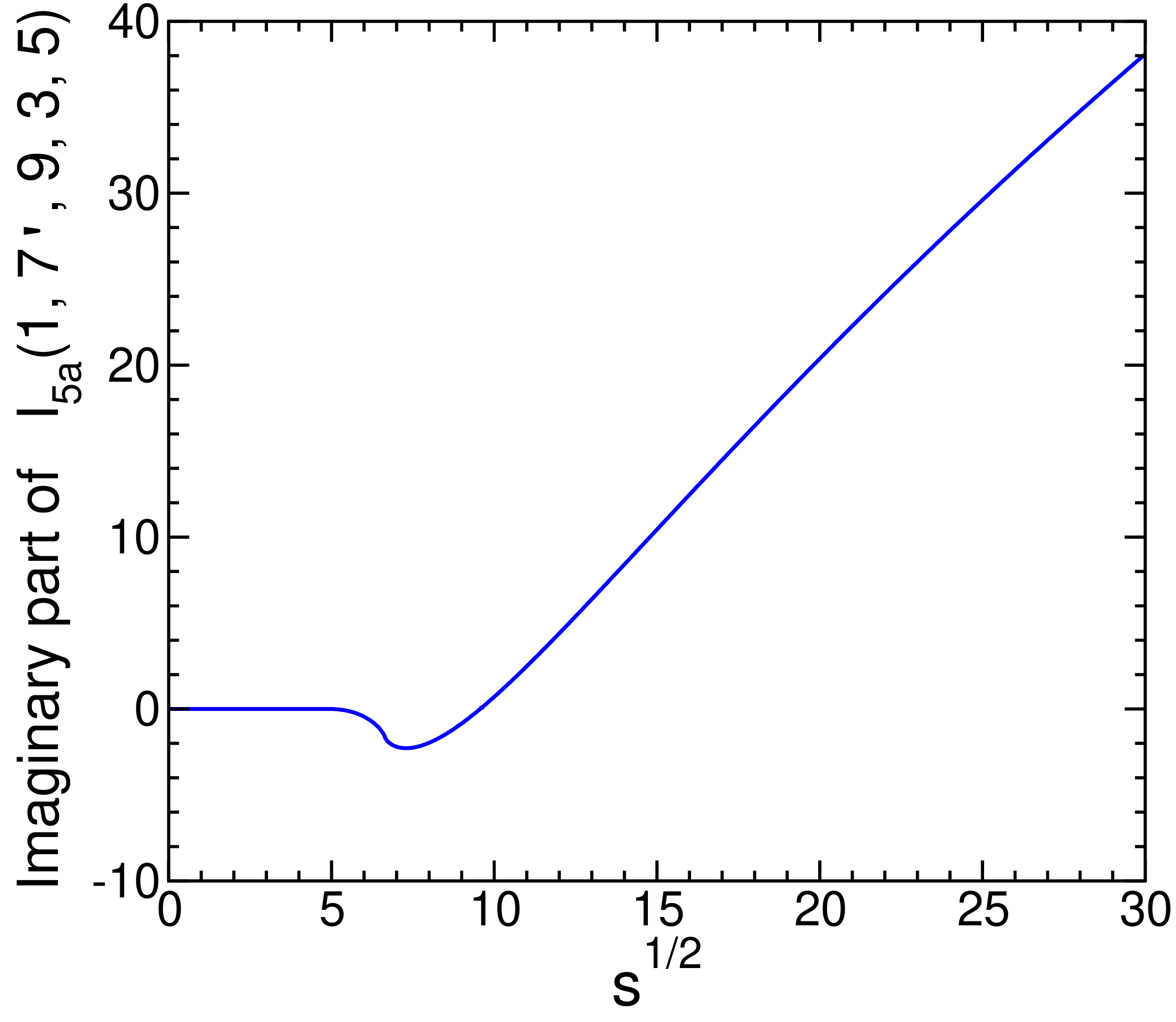}
\includegraphics[width=0.38\linewidth,angle=0]{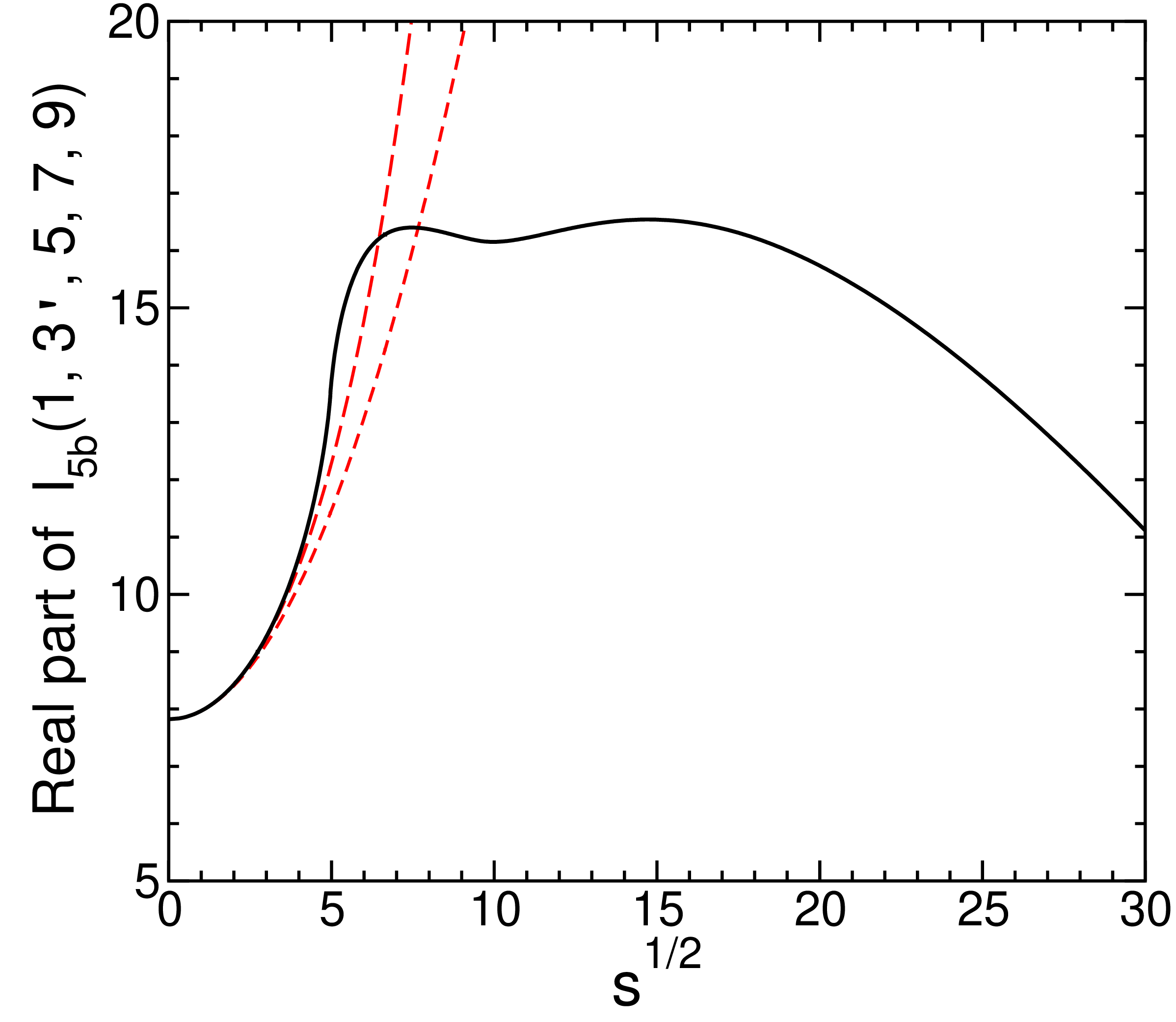}
\includegraphics[width=0.38\linewidth,angle=0]{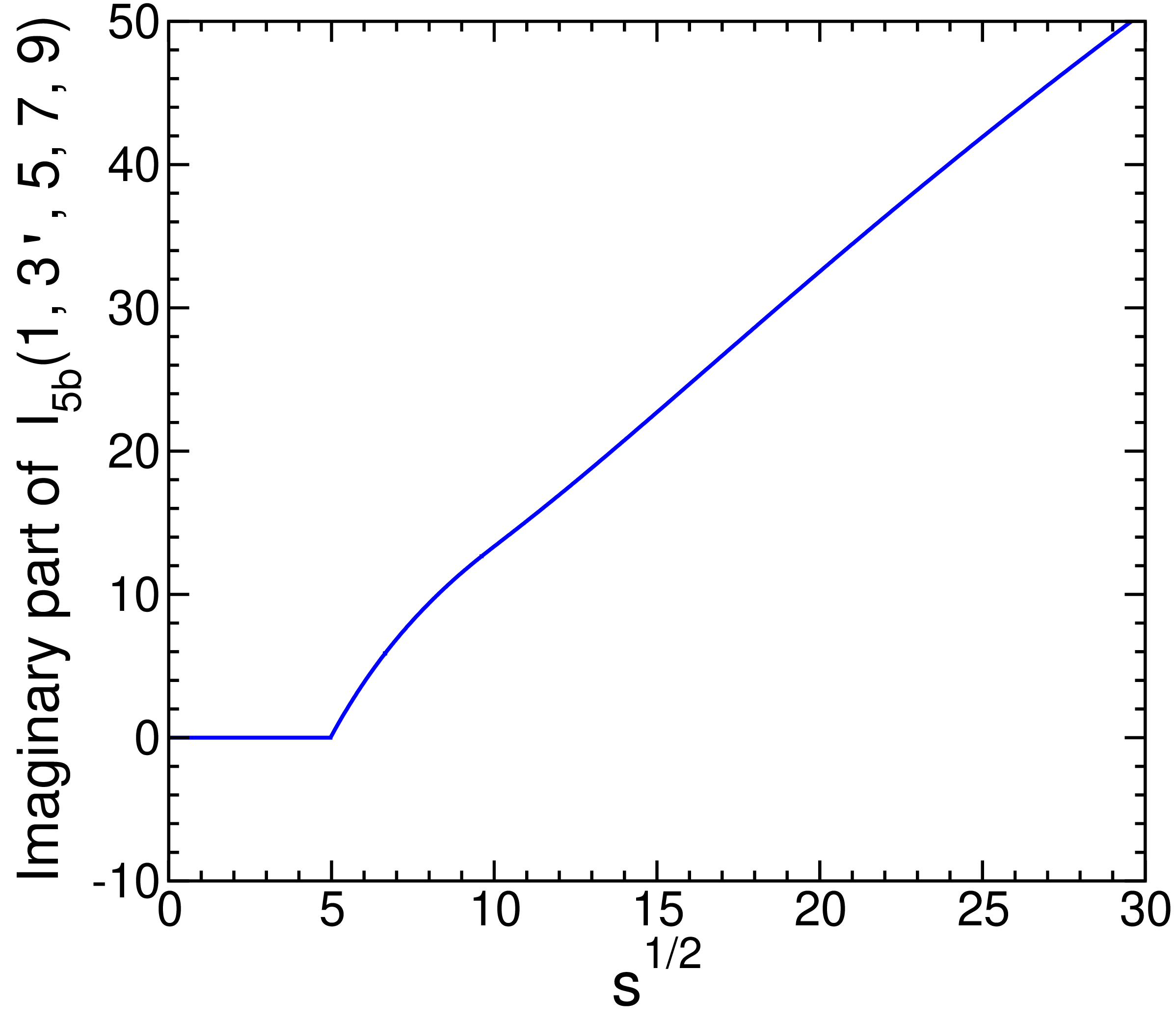}
\includegraphics[width=0.38\linewidth,angle=0]{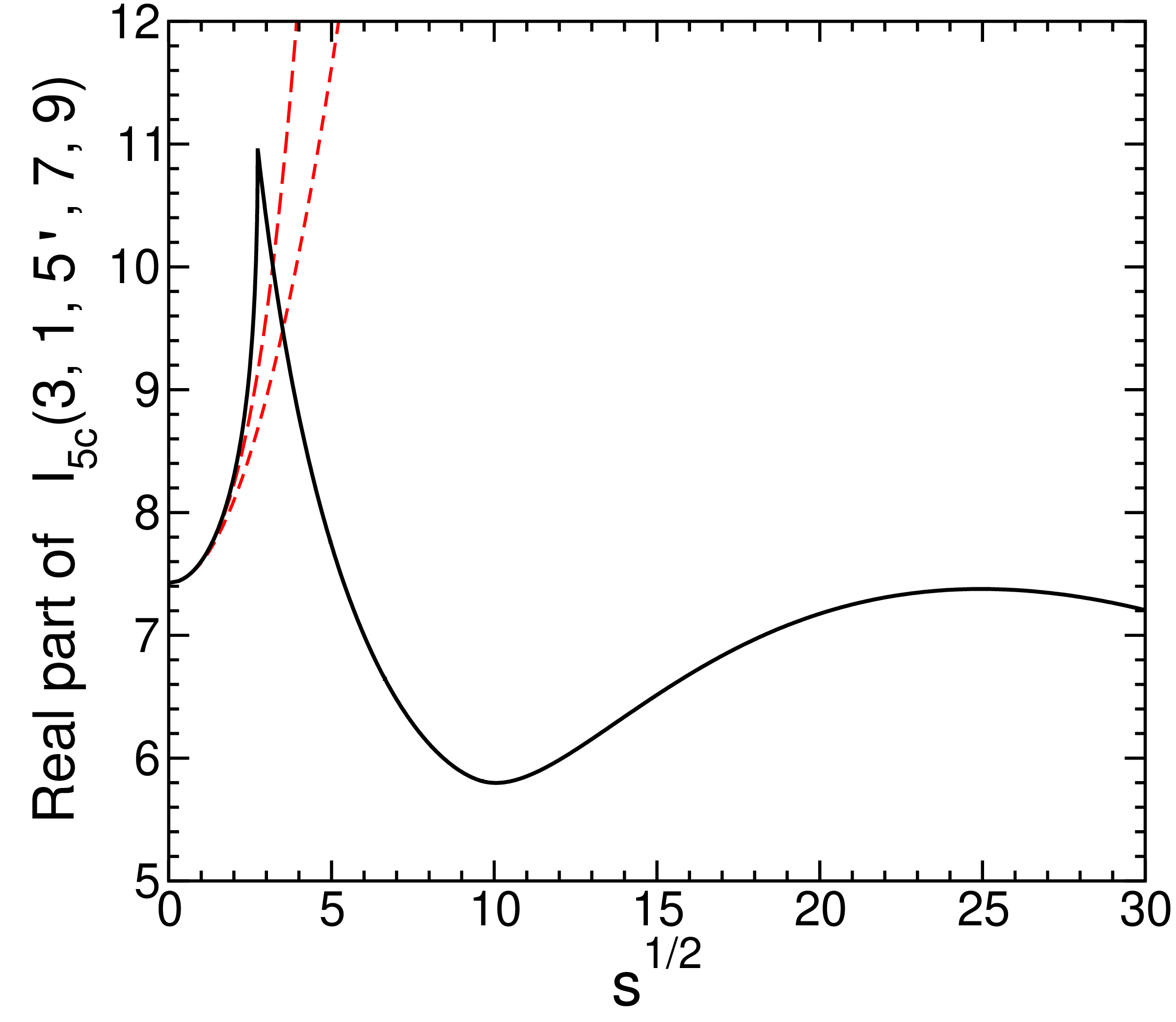}
\includegraphics[width=0.38\linewidth,angle=0]{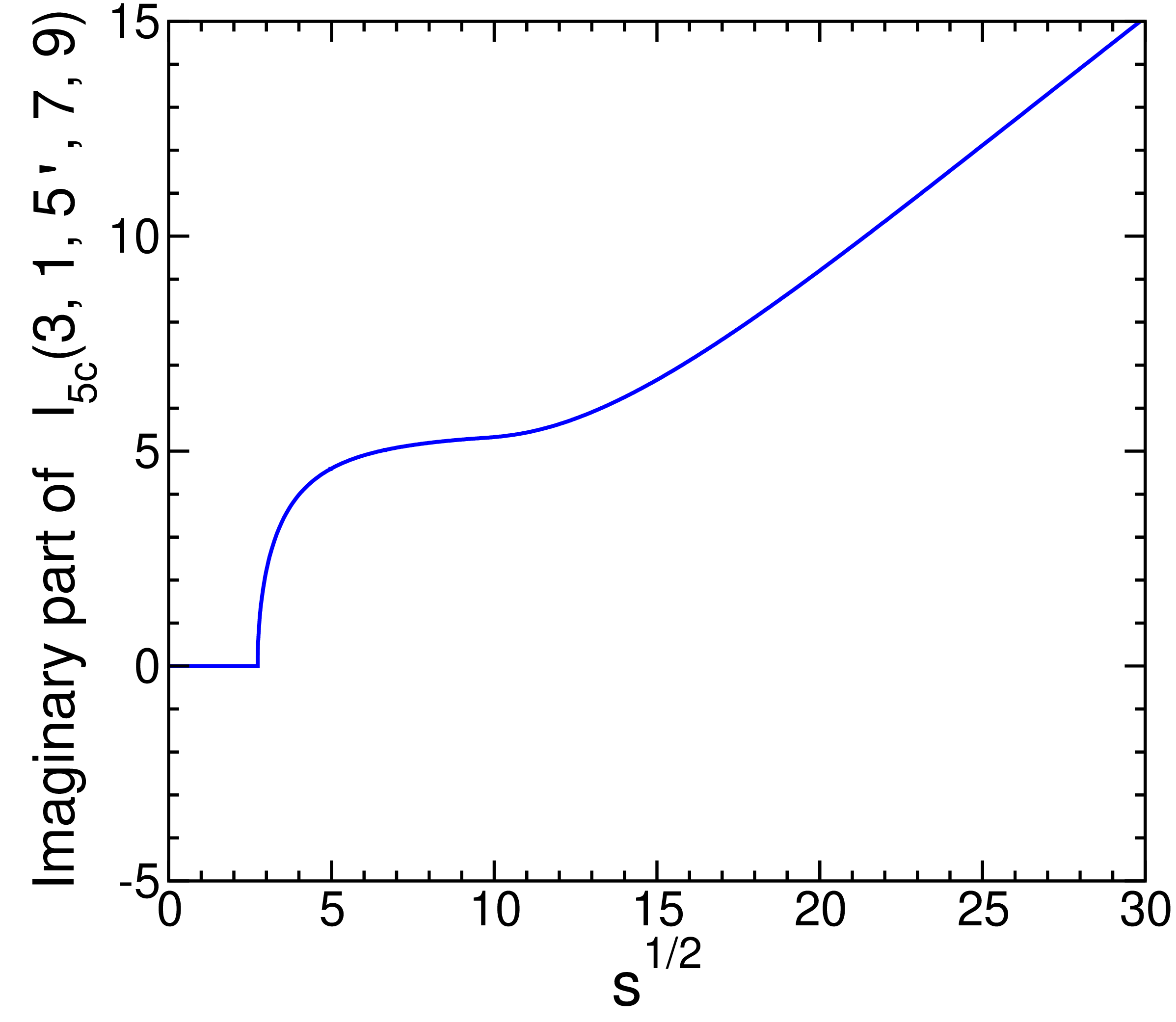}
\end{center}
\begin{minipage}[]{0.95\linewidth}
\caption{\label{fig:examples}Sample results, as a function of $\sqrt{s}$, for renormalized master integrals 
$I_4(3',5',7,9)$, $I_{5a}(1,7',9,3,5)$, $I_{5b}(1,3',5,7,9)$, and $I_{5c}(3,1,5',7,9)$. The left panels show the real parts, and the right panels show the imaginary parts. For the real part, the solid line is the full result, while the short-dashed and long-dashed lines are the expansions in small $s$ at order $s^1$ and $s^2$, respectively.
}
\end{minipage}
\end{figure}

As a first example, consider the master integrals for the topologies 
$I_4(3,5,7,9)$, $I_{5a}(1,3,5,7,9)$, $I_{5b}(1,3,5,7,9)$, and $I_{5c}(3,1,5,7,9)$. 
There is a 4-particle threshold at $\sqrt{s} = \sqrt{3} + \sqrt{5} + \sqrt{7} + 3 \approx 9.614$ for $I_4(3,5,7,9)$, $I_{5b}(1,3,5,7,9)$, $I_{5c}(3,1,5,7,9)$ and their derivatives; a 3-particle threshold at $\sqrt{s} = 1 + \sqrt{3} + \sqrt{5} \approx 4.968$ for $I_{5a}(1,7,9,3,5)$, $I_{5b}(1,3,5,7,9)$, and their derivatives;
another 3-particle threshold $\sqrt{s} = 4 + \sqrt{7} \approx 6.646$ for $I_{5a}(1,3,5,7,9)$ and its derivatives;
and a 2-particle threshold $\sqrt{s} = 1 + \sqrt{3} \approx 2.732$ (with cuspy behavior) for $I_{5c}(3,1,5,7,9)$ and its derivatives.
Results for four sample dimensionless master integrals are shown as a function of $\sqrt{s}$ in Figure \ref{fig:examples}, with real parts shown in the left panels and imaginary parts shown in the right panels. The imaginary parts turn on for $\sqrt{s}$ larger than the lowest threshold in each case.

\begin{figure}[!t]
\begin{center}
\includegraphics[width=0.45\linewidth,angle=0]{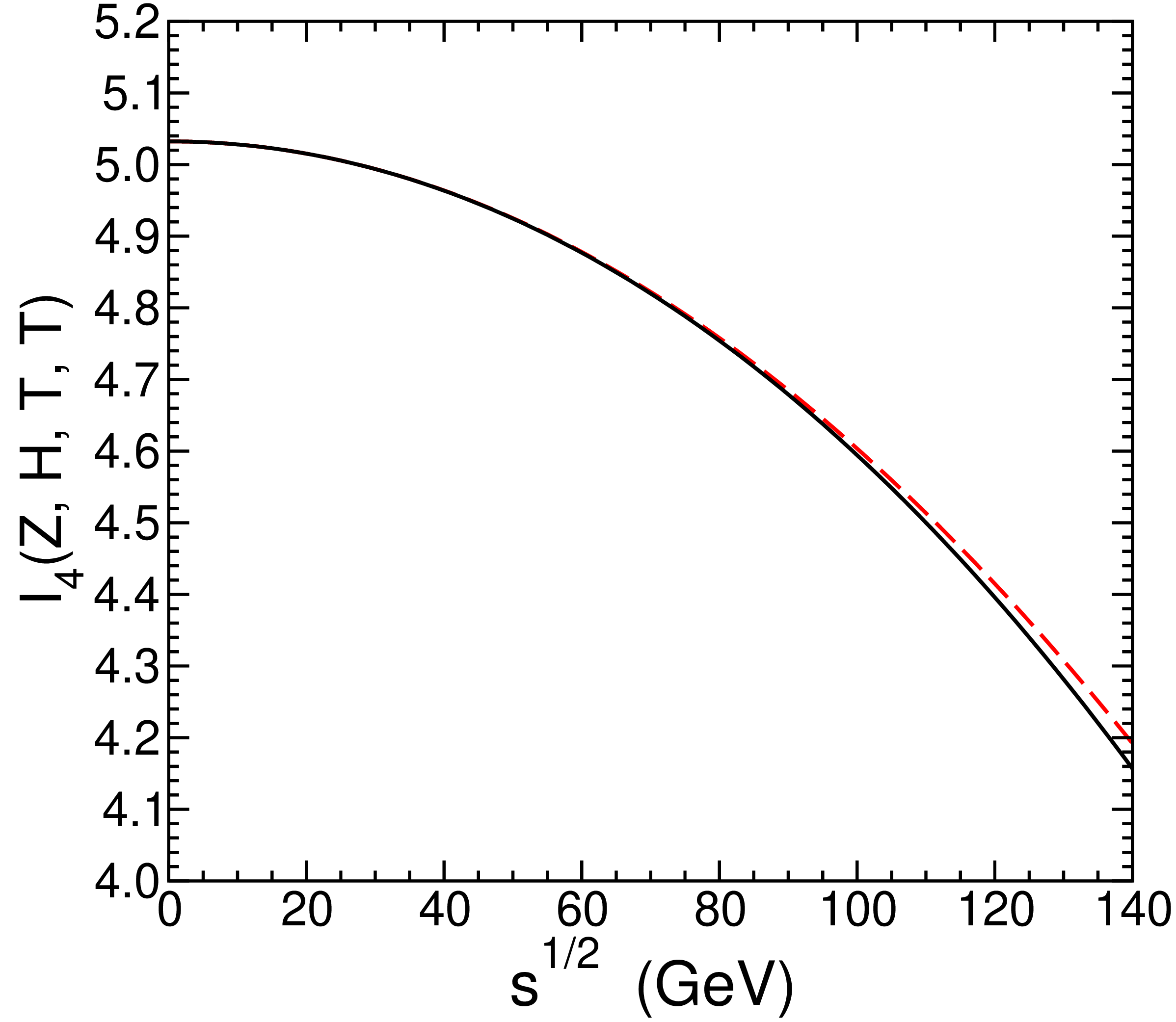}
\includegraphics[width=0.45\linewidth,angle=0]{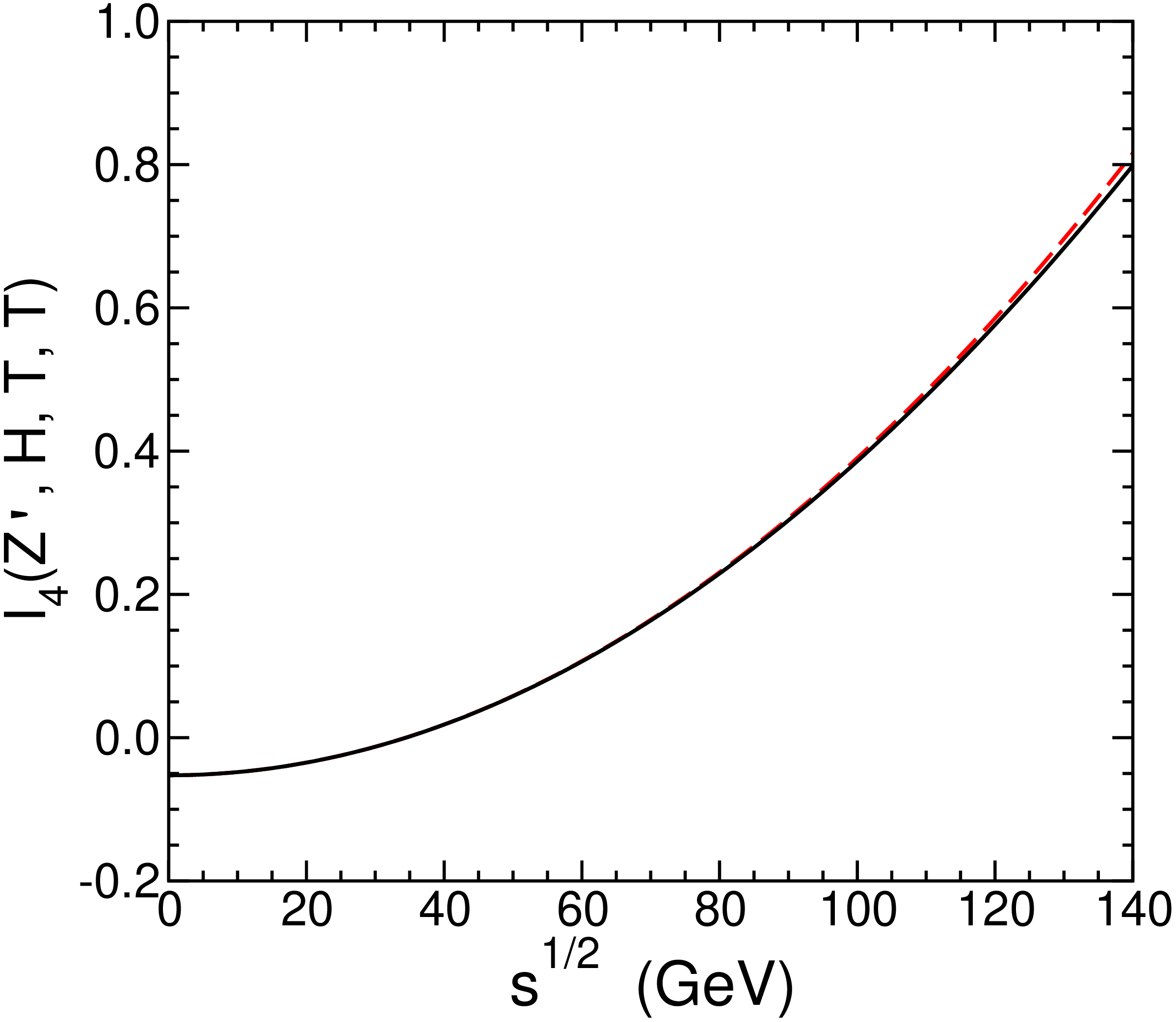}
\includegraphics[width=0.45\linewidth,angle=0]{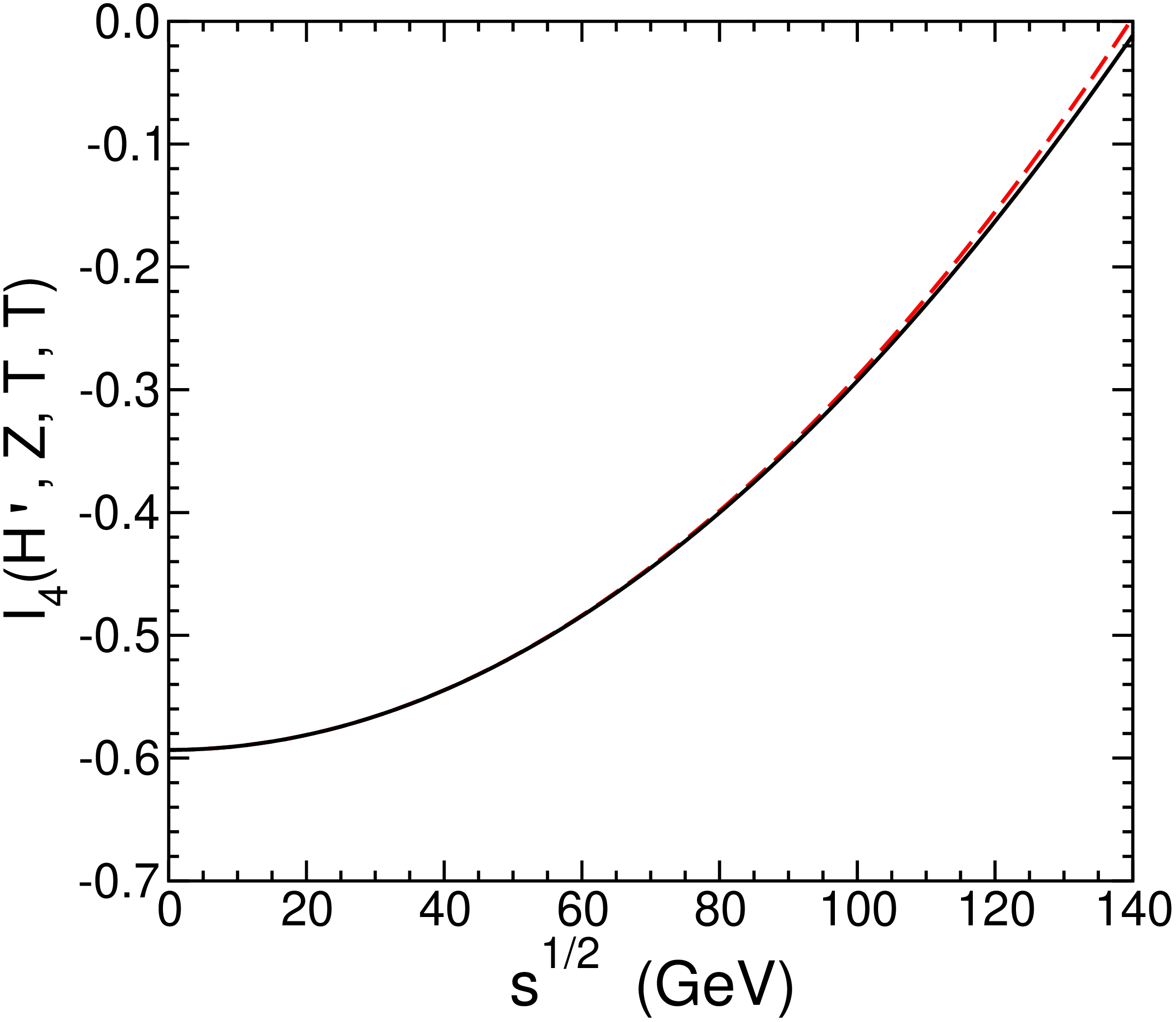}
\includegraphics[width=0.45\linewidth,angle=0]{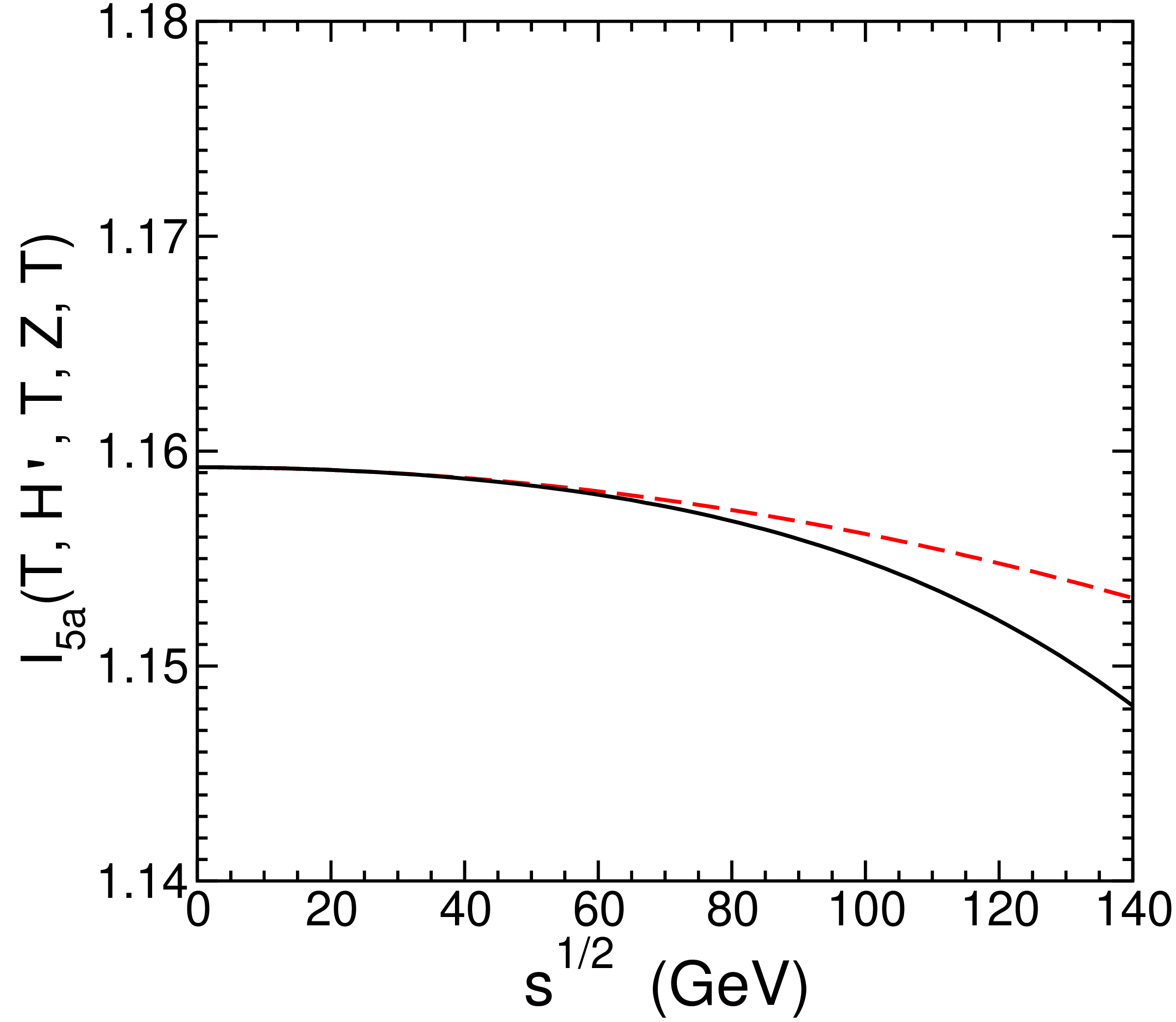}
\includegraphics[width=0.45\linewidth,angle=0]{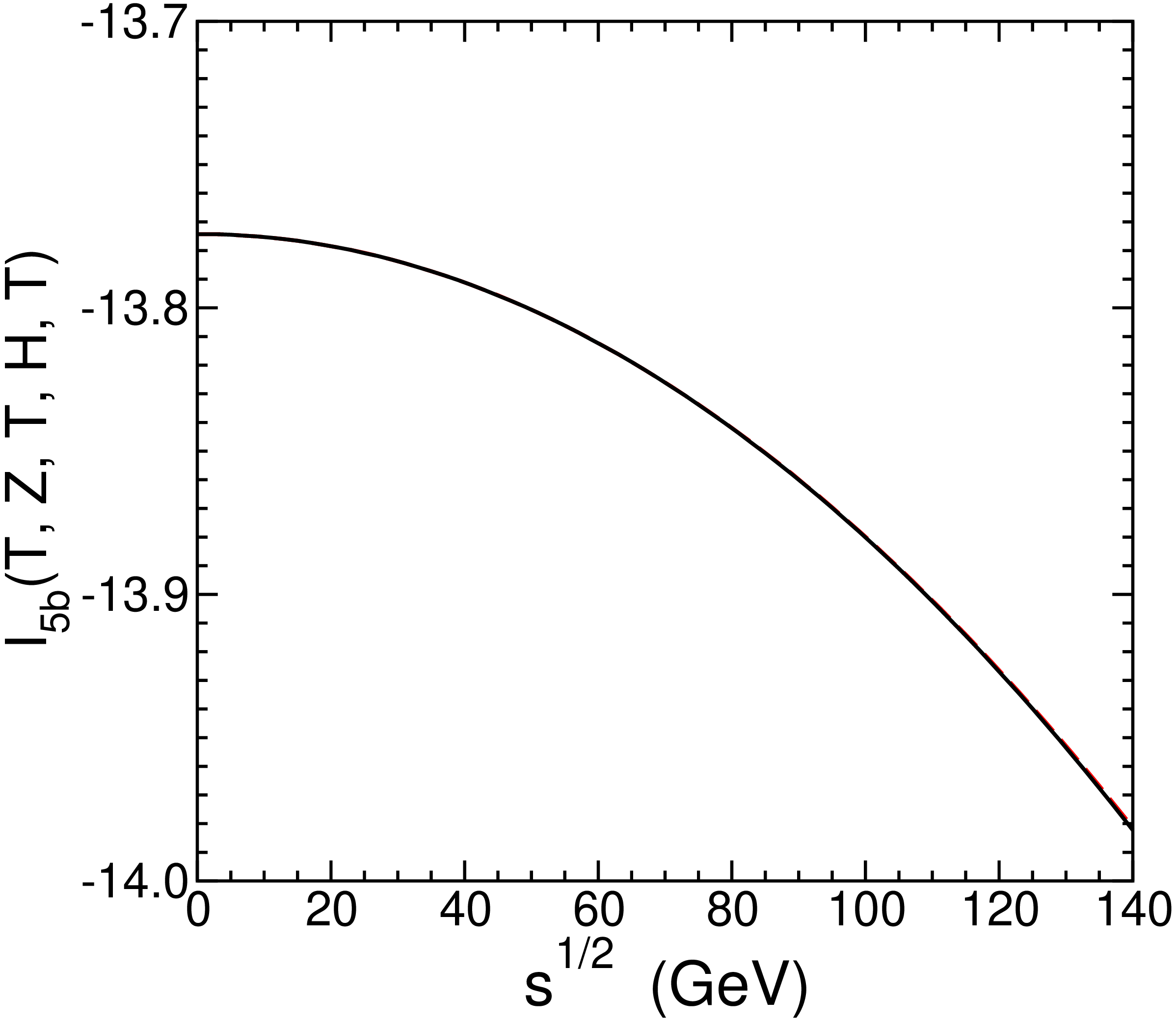}
\includegraphics[width=0.45\linewidth,angle=0]{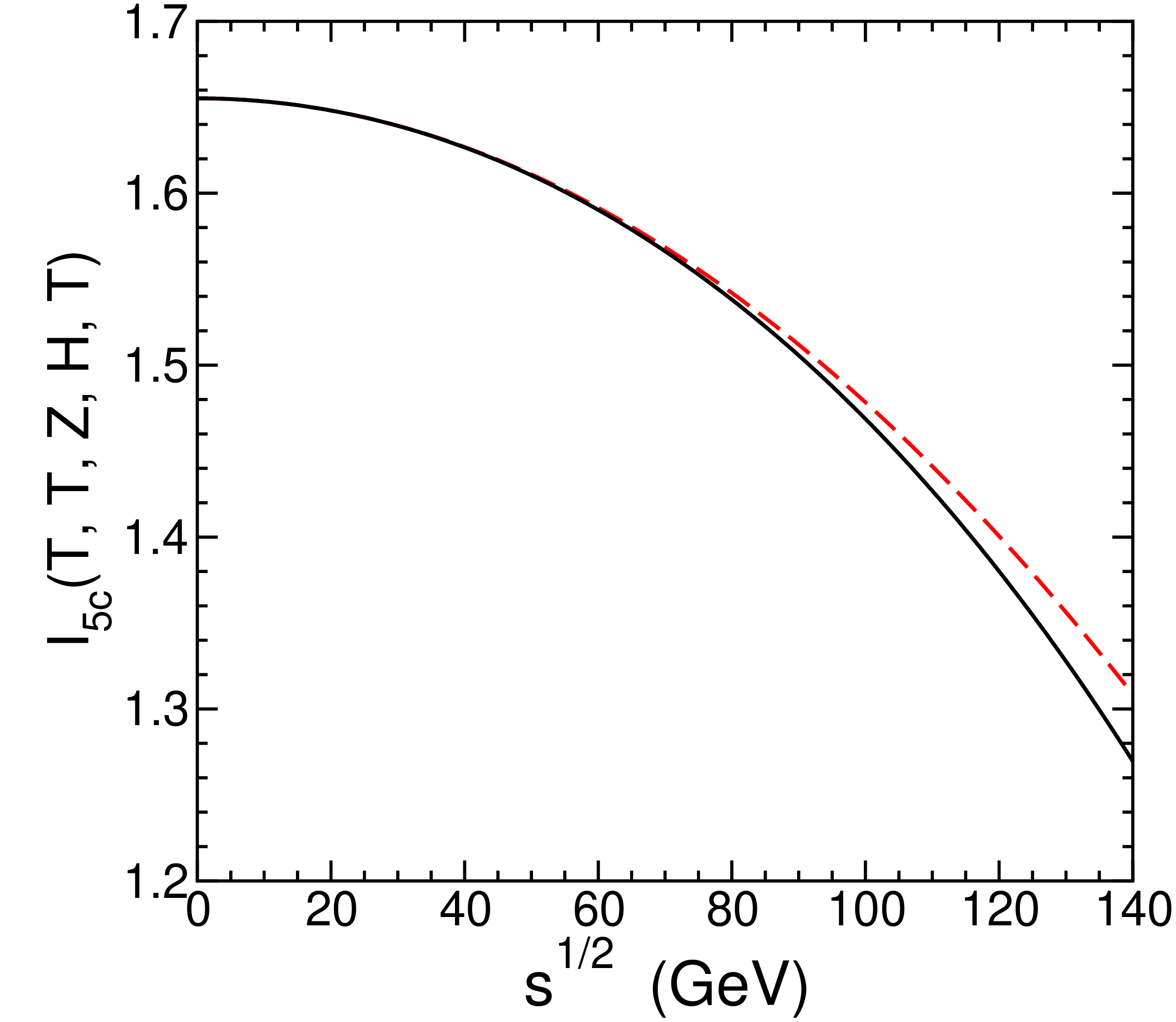}
\end{center}
\begin{minipage}[]{0.95\linewidth}
\caption{\label{fig:examplesSM}Results for renormalized master integrals 
$I_4(Z,H,T,T)$, $I_4(Z',H,T,T)$, $I_4(Z,H',T,T)$, $I_{5a}(T,H',T,Z,T)$, $I_{5b}(T,Z,T,H,T)$, and $I_{5c}(T,T,Z,H,T)$,
normalized in units of $T = \mbox{(173 GeV)}^2$, as a function of $\sqrt{s}$, for $H = \mbox{(125 GeV)}^2$ and $Z = \mbox{(91 GeV)}^2$. The solid line is the full result, while the dashed lines are the expansions in small $s$ at order $s^1$.
}
\end{minipage}
\end{figure}

As another benchmark example, relevant for the Standard Model, I consider the integrals obtained from the topologies $I_4(Z,H,T,T)$, $I_{5a}(T,H,T,Z,T)$, $I_{5b}(T,Z,T,H,T)$, and
$I_{5c}(T,T,Z,H,T)$, which arise in the 3-loop self-energies and pole masses of the Higgs and $Z$ bosons. For simplicity, I take squared mass arguments
\beq
T = Q &=& (\mbox{173 GeV})^2,
\\
H &=& (\mbox{125 GeV})^2,
\\
Z &=& (\mbox{91 GeV})^2,
\eeq
and present results using units in which the top-quark mass is 1, so that $T=1$, and 1 GeV $= 1/173$. (This is equivalent to multiplying each integral by the appropriate integer power of $T$ to make it dimensionless.) The results for some selected integrals are shown in Figure \ref{fig:examplesSM} for $\sqrt{s}$ up to 140 GeV. For comparison, the results of the expansions around $s=0$ up to linear order in $s$ are also shown. 
In these cases, the results of a series expansion to order $s^2$ would be visually almost indistinguishable from the full results on these plots. This makes it seem likely that simply expanding the integrals to order $s^2$ would be sufficient for practical results, at least for Higgs and $Z$ self-energy evaluation near the physical masses. However, some care is needed, because there could be cancellations between different master integrals in a given observable, and because in other mass configurations the small $s$ expansions of integrals
will not converge if there are lower thresholds. As benchmarks, the numerical results of all of the master integrals are given in Table \ref{tab:benchmarks} for $s = Z$ and $s=H$, to 16 digits of relative accuracy.

\begin{table}
\begin{center}
\begin{tabular}{| c | r | r |}
\hline
~Integral~ & ~$s=Z$~  &  ~$s=H$~
\\
\hline
\hline
$B(T,T)$ & 0.04744351586953098 & 0.09192546525780287 
\\[-2pt]
$S(Z, T, T)$ & $-$4.703771341470273 & $-$4.760582805362995
\\[-2pt]
$T(T, Z, T)$ & 0.08378683288496525 & 0.1364935723146822
\\[-2pt]
$T(Z, T, T)$ & $-$1.0837868328849654 & $-$1.0059164828526561
\\[-2pt]
$S(H, T, T)$ & $-$4.459767166902337 & $-$4.533014718203875 
\\[-2pt]
$T(T, H, T)$ & $-$0.1972725394703064 & $-$0.14936776548906433 
\\[-2pt]
$T(H, T, T)$ & $-$0.9092134235860295 & $-$0.8506322345109357
\\[-2pt]
$J_4(Z, H, T, T)$ & $-$3.7648277272371593 & $-$3.861531900214871
\\[-2pt]
$I_4(Z, H, T, T)$ & 4.671030470289084 & 4.340032890945725
\\[-2pt]
$I_4(T', Z, H, T)$ & $-$1.5389591085746437 & $-$1.4394297253491581
\\[-2pt]
$I_4(Z', H, T, T)$ & 0.31126684040808783 & 0.6287408011731227
\\[-2pt]
$I_4(H', Z, T, T)$ & $-$0.3439228747220906 & $-$0.1270716818364309
\\[-2pt]
$I_4(Z', T', H, T)$ & $-$2.392283625188141 & $-$2.2572720592690305
\\[-2pt]
$I_4(T', T', Z, H)$ & 0.5816634714095499 & 0.6783516671195731
\\[-2pt]
$I_4(H', T', Z, T)$ & $-$1.110806285033397 & $-$0.995579455324551
\\[-2pt]
$I_4(Z', H', T, T)$ & $-$5.479015505305125 & $-$5.317553339995855
\\[-2pt]
$I_{5a}(T, Z, T, H, T)$ & $-$13.622488723207809 & $-$13.488450608458654
\\[-2pt]
$I_{5a}(T', Z, T, H, T)$ & $-$2.0416981691719878 & $-$1.9715402873940417
\\[-2pt]
$I_{5a}(T, T', Z, H, T)$ & 2.5994642205657446 & 2.603265962001946
\\[-2pt]
$I_{5a}(T, Z', T, H, T)$ & 1.9742385083634955 & 1.9789726508216219
\\[-2pt]
$I_{5a}(T, T', H, Z, T)$ & 1.7041404263890743 & 1.7007051476320272 
\\[-2pt]
$I_{5a}(T, H', T, Z, T)$ & 1.1558139055810304 & 1.151248509704206 
\\[-2pt]
$I_{5b}(T, Z, T, H, T)$ & $-$13.86191527817819 & $-$13.939919181091156
\\[-2pt]
$I_{5b}(T', Z, T, H, T)$ & $-$2.096130262915311 & $-$2.0730906512207676 
\\[-2pt]
$I_{5b}(T, T', Z, H, T)$ & 2.622312357453426 & 2.6470911925905387
\\[-2pt]
$I_{5b}(T, Z', T, H, T)$  & 2.002565676860892 & 2.033414650054237
\\[-2pt]
$I_{5c}(T, T, Z, H, T)$ & 1.5021951295702383 & 1.354662946221542
\\[-2pt]
$I_{5c}(T, T, Z', H, T)$ & 2.59121942635416 & 2.634353749579119
\\[-2pt]
$I_{5c}(T, T, T', Z, H)$ & $-$0.4805187352659836 & $-$0.4883968583818474
\\[-2pt]
$I_{5c}(T, T, H', Z, T)$ & 1.2635192651839127 & 
1.2843120129525298
\\
\hline
\end{tabular}
\caption{Benchmark values for the renormalized master integrals following from the topologies $I_4(Z,H,T,T)$ and $I_{5a}(T,Z,T,H,T)$ and $I_{5b}(T,Z,T,H,T)$ and $I_{5c}(T,T,Z,H,T)$, for 
$T= Q = (\mbox{173 GeV})^2$,
$H= (\mbox{125 GeV})^2$,
and
$Z= (\mbox{91 GeV})^2$.
The results are given to 16 digits of relative accuracy, and in units such that the top-quark mass is unity, so that $T=1$ and 1 GeV $= 1/173$. This is equivalent to multiplying
each integral by the appropriate power of $T$ to make it dimensionless.
\label{tab:benchmarks}}
\end{center}
\end{table}
\newpage

\section{Outlook\label{sec:outlook}}
\setcounter{equation}{0}
\setcounter{figure}{0}
\setcounter{table}{0} 
\setcounter{footnote}{1}

In this paper, I have provided results for the master integrals for 3-loop self-energy integrals with 4 or 5 propagators with generic masses. Provided in ancillary files in computer readable form, these results include the derivatives with respect to each of the squared masses and the external momentum invariant. In particular, the results for derivatives with respect to $s$ enable the numerical computation of the renormalized master integrals for general arguments, using the coupled first-order differential equations starting from (or near) $s=0$ and integrating along a contour in the upper half complex $s$ plane. 

In some cases of non-generic masses that are either equal to each other or to 0, the results as given above require some care, because the polynomials in denominators of some of the identities can vanish identically for all $s$, for example due to the appearance of $\Psi(x,x,y,y) = 0$ or $\Delta(0,x,x) = 0$. The corresponding identities between master integrals, and elimination of non-masters, can be derived either by reprising the procedure outlined in this paper with the non-generic mass relations implemented, or simply by taking limits of the identities given here when put into polynomial coefficient form. In the cases considered in the present paper, the offending denominators do not appear in the derivatives with respect to $s$ anyway, so that there is no obstacle to their numerical computation. 
In particular, there are no Standard Model master integrals with 4 and 5 propagators for which the limits cannot be obtained very simply, except the ones with 0 masses already covered in ref.~\cite{Martin:2021pnd} and references therein.

For practical applications, it will be necessary to extend these results to the remaining 3-loop self-energy master integrals with 6, 7, and 8 propagators, since self-energies and pole masses of scalars, fermions, and vector bosons in the Standard Model and its extensions will always involve such integrals. I think it is likely that a relatively efficient way to obtain those results will be to use the same sort of approach as in this paper, relying on the form guaranteed by the structure of the IBP relations but without actually following the IBP reduction and elimination procedure. 
It would be interesting to see whether traditional IBP methods and codes can produce the results for general masses.
In any case, the eventual goal will be to produce computer code that can evaluate all pertinent renormalized master integrals for a given 3-loop self-energy topology on demand, and an algorithm that can reduce any given self-energy loop integral functions, including those involving non-trivial numerators, to the masters. The latter algorithm might be applied only at the numerical level (perhaps in terms of rational numbers that closely approximate physical masses), because of the extreme algebraic complexity involved if the squared masses are general and treated symbolically.

The expansion method outlined in section \ref{sec:expansions} can be applied in the very same way to the topologies that were called $I_{6a}$, $I_{6c}$, $I_{6d}$, and $I_{7d}$ in Figure 3.2 of ref.~\cite{Martin:2021pnd}. The expansions of $s$ for the remaining diagram topologies with 6, 7, or 8 propagators will not be quite so straightforward, since they are not of the form assumed in section \ref{sec:expansions}. However, they can in principle always be found by simply expanding denominators to move all $p^\mu$ factors to numerators, resulting in linear combinations of scalar vacuum integrals, which can in turn be reduced to masters. More optimistically, it also seems plausible to me that one can instead obtain more general all-orders formulas for the expansions in $s$, in terms of differential operators containing derivatives of the masses acting on the vacuum integrals, similar to and generalizing eqs.~(\ref{eq:sexpansion})-(\ref{eq:definean}).  

\newpage
 
{\it Acknowledgments:} This work was supported in part by the National Science 
Foundation grant number 2013340. 


\end{document}